\documentclass[review]{elsarticle}

\usepackage{url}
\usepackage{amssymb}
\def\doi#1{\href{https://doi.org/\detokenize{#1}}{\url{https://doi.org/\detokenize{#1}}}}

\usepackage{graphicx}
\usepackage{wrapfig}
\usepackage{soul}
\usepackage{color}
\usepackage{subcaption}
\usepackage{float}

\usepackage{listings}
\usepackage{diagbox}
\usepackage{algorithm}
\usepackage{algpseudocode}
\usepackage{xcolor}
\usepackage{textcomp}
\usepackage{paralist}

\definecolor{codegreen}{rgb}{0,0.6,0}
\definecolor{codegray}{rgb}{0.5,0.5,0.5}
\definecolor{codepurple}{HTML}{C42043}
\definecolor{codeblue}{HTML}{0000FF}
\definecolor{backcolour}{HTML}{FFFFFF}
\definecolor{bookColor}{cmyk}{0,0,0,0.90}  
\color{bookColor}

\lstdefinestyle{mystyle}{
	backgroundcolor=\color{backcolour},   
	commentstyle=\color{codegreen},
	keywordstyle=\color{codeblue},
	numberstyle=\numberstyle,
	stringstyle=\color{codepurple},
	basicstyle=\linespread{1}\scriptsize\upshape\ttfamily,
	breakatwhitespace=true,
	breakautoindent=true,
	breaklines=true,
	breakindent=0pt,
	captionpos=t,
	keepspaces=true,
	numbers=left,
	numbersep=5pt,
	showspaces=false,
	showstringspaces=false,
	showtabs=false,
	tabsize=2,
	showlines=true,
        escapeinside=||,
	framesep=1pt,
	xleftmargin=18pt,
	framexleftmargin=18pt,
	frame=tb,
	framerule=1pt,
        breaklines=true                    
}
\newcommand\numberstyle[1]{%
	\scriptsize	
	\color{codegray}%
	\ttfamily
	\ifnum#1<10 \fi#1|%
}

\newcommand \UntilLTL      {\mathbin{\mathcal{U}\kern-.1em}}

\usepackage{xcolor,colortbl}

\definecolor{Gray}{gray}{0.85} 
\definecolor{Grey}{HTML}{EFEFEF}
\definecolor{LightCyan}{rgb}{0.88,1,1}
\definecolor{grannysmithapple}{rgb}{0.66, 0.89, 0.63}
\definecolor{honeydew}{rgb}{0.94, 1.0, 0.94}
\definecolor{lavenderblush}{rgb}{1.0, 0.94, 0.96} 
\definecolor{lightblue}{rgb}{0.68, 0.85, 0.9}
\definecolor{lightapricot}{rgb}{0.99, 0.84, 0.69}

\newcolumntype{o}{>{\columncolor{Grey}}c}
\newcolumntype{x}{>{\columncolor{LightCyan}}c}

\newtheorem{definition}{Definition}
\usepackage{tikz}
\usetikzlibrary{decorations.pathreplacing}
\usetikzlibrary{calc}

\newcommand{\tikzmark}[1]{\tikz[remember picture,overlay]\node[yshift=1pt](#1){};}

\usepackage{array,multirow}

\journal{Information Systems}

\begin{document}

	\begin{frontmatter}

\def\RunningHead{{Timed Ordered Anti-Patterns over Graph-encoded Event Logs}}

\title{Validating Temporal Compliance Patterns: A Unified Approach with $MTL_{f}$ over various Data Models}

   \author[mymainaddress]{Nesma M. Zaki\corref{mycorrespondingauthor}}
		\cortext[mycorrespondingauthor]{Corresponding author}
		\ead{n.zaki@fci-cu.edu.eg}
   \author[mymainaddress]{Iman M. A. Helal}
		\ead{i.helal@fci-cu.edu.eg}
    \author[mymainaddress]{Ehab E. Hassanein}
        \ead{e.ezat@fci-cu.edu.eg}
    \author[mymainaddress,add2]{Ahmed Awad}
		\ead{ahmed.awad@buid.ac.ae}
		
    \address[mymainaddress]{Faculty of Computers and Artificial Intelligence, Cairo University, Egypt}
    \address[add2]{The British University in Dubai, Dubai, The UAE}
      

\begin{abstract} 
Process mining extracts valuable insights from event data to help organizations improve their business processes, which is essential for their growth and success. By leveraging process mining techniques, organizations gain a comprehensive understanding of their processes' execution, enabling the discovery of process models, detection of deviations, identification of bottlenecks, and assessment of performance. Compliance checking, a specific area within conformance checking, ensures that the organizational activities adhere to prescribed process models and regulations. Linear Temporal Logic over finite traces ($LTL_{f}$ ) is commonly used for conformance checking, but it may not capture all temporal aspects accurately. This paper proposes Metric Temporal Logic over finite traces ($MTL_{f}$ ) to define explicit time-related constraints effectively in addition to the implicit time-ordering covered by $LTL_f$. Therefore, it provides a universal formal approach to capture compliance rules. Moreover, we define a minimal set of generic $MTL_f$ formulas and show that they are capable of capturing all the common patterns for compliance rules.

As compliance validation is largely driven by the data model used to represent the event logs, we provide a mapping from $MTL_f$ to the common data models we found in the literature to encode event logs, namely, the relational and the graph models. A comprehensive study comparing various data models and an empirical evaluation across real-life event logs demonstrates the effectiveness of the proposed approach.


\end{abstract}

	
\end{frontmatter}

\section{Introduction}\label{sec:introduction}

Improving business processes is crucial for achieving organizational success. Process mining analyzes event logs of organizations’ information systems to provide actionable insights that can improve business processes. By leveraging process mining techniques~\cite{processMiningBook2016}, organizations gain a comprehensive understanding of how their processes are executed in reality. By analyzing event logs, organizations can make decisions to enhance process efficiency, reduce lead times, and improve compliance. Conformance checking~\cite{conformanceCheckingBook18}, in particular, helps organizations to identify areas for improvement, enhance compliance, and drive improvements to ensure processes operate properly. 

Compliance checking~\cite{complianceRequirements2019} is a specialization of conformance checking. It focuses on verifying whether organizational activities align with established regulations, standards, laws, or internal policies~\cite{complianceRequirements2019}. These rules serve to restrict the behaviour of processes concerning control flow, data, resources, and timing. The goal of compliance checking is to identify process instances that violate the rules. There have been several proposals to model business process compliance rules. Most proposals provide a visual representation that is very close to how business processes are modelled, e.g.,~\cite{DECLARE07,DBLP:conf/icsoc/AwadWW09,DBLP:journals/corr/abs-1110-4161,DBLP:journals/emisa/KnupleschRLKR14}. Nevertheless, they largely utilise temporal logic as a formal basis underpinning their notation and as a means to analyse the behaviour of the business processes. In the BPM community, it is commonly accepted to use, $LTL_f$, linear temporal logic on finite traces, to formalise compliance patterns~\cite{dwyer1999patterns, elgammal2016formalizing}, define declarative process models~\cite{DECLARE07}, or conduct conformance checking of declarative business processes~\cite{maggi2012runtime}. For example, in an order fulfilment process, the order has to be paid before the delivery is made. This can be easily captured in $LTL_f$\footnote{We use $LTL$ and $LTL_f$ interchangeably}. However, $LTL$ is expressively limited when it comes to modelling explicit time constraints in compliance requirements~\cite{DBLP:journals/emisa/KnupleschRLKR14,surveyComplianceChecking18}. For instance, in the same delivery process, we might have a requirement that the order has to be paid within \emph{24 hours} from order placement; otherwise, it will be cancelled. In the former example, the time-ordering constraint was implicit. In the latter, it is explicit with.

In this paper, we argue that $LTL_f$ is insufficient to capture compliance rules that require explicit time constraints. We propose to use the metric time logic (MTL) as a more expressive formalism that is capable of capturing both implicit and explicit time-ordering rules. More specifically, we utilise $MTL_f$~\cite{de2021timed} for finite traces, as this fits the nature of compliance checking over event logs. Moreover, we provide a minimal set of compliance patterns that cover the common patterns for compliance rules and declarative conformance checking. The minimal set generalizes the common $LTL_f$ patterns with explicit time constraints.

The validation of compliance rules, instances of compliance patterns, against event logs, is usually reduced to a model-checking problem, where logs are transformed into a labelled transition, an automaton, fitting the temporal logic used~\cite{DBLP:conf/bpm/GrandoAM11}, or an alignment of the log with the automaton~\cite{de2012aligning,burattin2012techniques,DBLP:journals/is/LeoniMA15}. However, the construction time of the automaton is exponential to the size of $LTL$ formulas~\cite{DBLP:journals/tweb/MontaliPACMS10}.

To enable efficient conformance checking, several approaches were proposed to encode event logs using various data models, i.e., relational data model~\cite{declarativeMiningSQL16} or graph data model~\cite{multiDimlEventGrahDB21}. Selecting appropriate storage methods becomes crucial to ensure that the compliance checking process remains responsive, even when dealing with a large number of traces. The encoding of event logs in a certain data model entails the translation of compliance rules from the temporal logic formal specification to the query language of the data model, e.g.~\cite{declarativeMiningSQL16,relationalPMOperator20,DBLP:conf/bpm/RivaBMMM23} for the relational encoding, and ~\cite{multiDimlEventGrahDB21,Antipatterns} in the case of graph databases encoding. In this paper, we contribute translation algorithms from $MTL_f$ formulas to the common event log encoding models that we found in the literature. Moreover, we conduct an empirical evaluation across the different encodings and compliance rule representations to assess the expressiveness and scalability of the encoding as log sizes increase. In summary, we make the following contributions:

\begin{itemize}
        \item Propose the use of $MTL_f$ as the formal basis for compliance rules, both with implicit and explicit time constraints,
        \item Establish a concise formulation for the common compliance patterns,
        \item Develop mappings from the $MTL_f$ concise formulas to the different query languages, namely, $SQL$, $Match\_Recognize$, and $Cypher$,
         \item Conduct an empirical evaluation to compare the results with graph representation, relational data models, and Linear Temporal Logic with Declare Analyzer. The comparison is conducted across four real-life event logs.
\end{itemize}



The rest of this paper is organized as follows: Section~\ref{sec:background} provides an overview of the background concepts and techniques that are used throughout the paper. Related work is discussed in Section
\ref{sec:related:work}. Section~\ref{sec:formalization} presents how \emph{$MTL_{f}$} can be used to formalize compliance patterns. The translation from $MTL_f$-based compliance patterns to the different data models is presented in Section~\ref{sec:proposed:approach}. The empirical evaluation of the data different event log encoding for compliance checking is detailed in Section \ref{sec:eval}, before we conclude the paper with an outlook on future work in Section~\ref{sec:conclusion:future}.

\section{Preliminaries}\label{sec:background}

This section lays the background concepts necessary to follow the content of the paper. Section~\ref{sub:sec:background:definitions} formalizes the definitions related to events representation. Section~\ref{sub:sec:temp:logic} introduce temporal logic: $LTL$, $LTL_f$, and $MTL_f$. Section~\ref{sub:sec:compliance:patterns} reviews the conformance and compliance patterns common for process mining. Finally, Section~\ref{sec:running:example} illustrates an exemplary business process and a set of compliance requirements that will be used to illustrate the contributions in the rest of the paper.

\subsection{Events, Traces, and Logs}\label{sub:sec:background:definitions}


\begin{definition}[Event]\label{def:event} An event $e$ is a tuple $(a_1,a_2,\dots, a_n)$ where $a_i$ is an attribute value drawn from a respective domain $a_i \in D_i$. At least three domains and their respective values must be defined for each event $e$: $D_c$, the set of case identifiers, $D_a$, the set of activity identifiers, and $D_t$, the set of timestamps. We denote these properties as $e.c$, $e.a$, and $e.t$ respectively. Other properties and domains are optional such as $D_r$, the resources who perform the tasks, $D_l$, the lifecycle phase of the activity.\end{definition}

We reserve the first three properties in the event tuple to reflect the case, the activity label, and the timestamp properties.

\begin{definition}[Trace]\label{def:trace} A trace is a finite sequence of events $\sigma = \langle e_1, e_2,\dots, e_m\rangle$ where $e_i$ is an event, $1 \leq i \leq m$ is a unique position for the event that identifies the event $e_i$ in $\sigma$ and explicitly positions it, and for any $e_i, e_j \in \sigma: e_i.c = e_j.c$
\end{definition}

\begin{definition}[Event log]\label{def:event:log} An event log is a finite sequence of events $\mathcal{L} = \langle e_1, e_2,\dots, e_m\rangle$ where events are ordered by their timestamps for any $e_i$ and $e_{i+1}: e_i.t \leq e_{i+1}.t$.
\end{definition}

\subsection{Temporal Logic} \label{sub:sec:temp:logic}

Temporal Logic has been widely utilized for specification and verification of reactive systems, as well as to validate compliance with regulations and constraints in a variety of fields. Linear Temporal Logic (LTL) and Metric Temporal Logic (MTL) are two specific temporal logic sub-types with different focuses and capabilities.

\begin{definition}[Linear-Time Temporal Logic (LTL)] $LTL$~\cite{rozier2011linear} is an extension of propositional Boolean logic that includes modalities for expressing temporal operators. LTL formulas are composed of a set of propositional variables $P$, temporal operators such as $\UntilLTL$ (Until), $\square$ (also noted as \emph{G} for globally), $\square^{-1}$ (also noted as \emph{G} for globally in the past), $\Diamond$ (means \emph{eventually}), $\Diamond^{-1}$ (means \emph{eventually in the past}), $\bigcirc$ (means \emph{next state}) and $\bigcirc^{-1}$ (means \emph{previous state}).
\end{definition}

 The formal definition of the collection of LTL formulas over $P$ is as follows:
 \begin{itemize}
\item Each $p \in P$ is an LTL formula.
 \item If $\phi,\psi$ are LTL formulas then \[\ \phi,\psi  := \psi \UntilLTL \phi~|~\bigcirc (\phi)~|~\bigcirc^{-1}(\phi)~|~\phi \vee \psi~|~\square \phi~|~\square^{-1} \phi~|~\Diamond \phi~|~\Diamond^{-1} \phi\] are LTL formulas. 
 \end{itemize}

Linear Temporal Logic on finite traces (\emph{$LTL_{f}$}) \cite{li2014ltlf,fionda2018ltl,LTLf} is a variant of $LTL$ that restricts $LTL$ formula semantics to finite traces. It allows for expressing temporal properties over sequences of events, where each event occurs exactly once. $LTL_{f}$ formulas are evaluated on finite traces, which are partial orders of events. They are used in the context of reactive synthesis and runtime verification. $LTL_f$ is widely used to formalize declarative business process models~\cite{maggi2012runtime}, declarative conformance checking~\cite{chiariello2022asp}, and compliance checking.

\begin{definition}[Metric Temporal Logic on finite traces ($MTL_{f}$)]\label{def:MTL} $MTL_{f}$ is a generalization of $LTL_{f}$ in which temporal operators are substituted by time-constrained counterparts, such as until~\emph{($\UntilLTL_{I}$)}, next~\emph{($\bigcirc_{I}$)}, previous~\emph{($\bigcirc^{-1}_{I}$)}, globally ~\emph{($\square_{I}$)}, globally in the past ~\emph{($\square^{-1}_{I}$)}, eventually ~\emph{($\Diamond_{I}$)},and eventually in the past ~\emph{($\Diamond^{-1}_{I}$)} operators, where I $\subseteq [0,\infty)$ represents an interval of real numbers with endpoints in the set of natural numbers (N) till $\infty$. 

\end{definition}

The construction of $MTL_{f}$~\cite{de2021timed} formulas is similar to that of $LTL_{f}$, with the exception that temporal operators are now constrained by an interval \emph{I}. $MTL_{f}$ well-formed equations are formed following the rule~\cite{thati2005monitoring}: 
\[\phi,\psi:= p~|~\neg\phi~|~(\phi \wedge \psi)~|~\bigcirc_I \phi~|~(\phi \UntilLTL_I \psi)~|~\square_{I} \phi~|~\square^{-1}_{I} \phi~|~\Diamond_{I} \phi~|~\Diamond^{-1}_{I} \phi\] This expression $\phi \UntilLTL_I \psi$ defines behavior modeled expressed as timed words containing a sequence of $\phi$ followed by $\psi$ which occurs at within time interval I. For example, $\phi \UntilLTL_{\geq 3} \psi$ represents $\phi \UntilLTL_{[3,\infty)} \psi$, $\phi \UntilLTL_{\leq 3} \psi$ represents $\phi \UntilLTL_{[0,3])} \psi$, $\phi \UntilLTL_{3} \psi$ represents $\phi \UntilLTL_{[3,3]} \psi$ and $\phi \UntilLTL_{2 \leq i \leq 3} \psi$ represents $\phi \UntilLTL_{[2,3]} \psi$. This notation allows for specifying different time intervals and constraints between $\phi$ and $\psi$. It is important to note that Linear Temporal Logic ($LTL_{f}$) is a subset of $MTL_{f}$, where only the interval $[0, \infty)$ is permitted.


\subsection{Compliance Patterns}\label{sub:sec:compliance:patterns}

Conformance checking~\cite{conformanceCheckingBook18} 
provides an alignment between a trace and a
process model to quantify the deviation between the required behavior (the model) and the observed behavior (the trace). In many situations, it is necessary to check deviations at a finer granularity, such as the level of activities, e.g., absence, existence, or pairs of activities, e.g., coexistence, mutual exclusion, along with time window constraints. Such finer granularity checks are referred to as compliance checking, and \emph{compliance patterns} are used for categorizing the types of compliance requirements~\cite{complianceRequirements2019}. 

\begin{figure}[ptb!]
\centering

\includegraphics[width=1\linewidth]{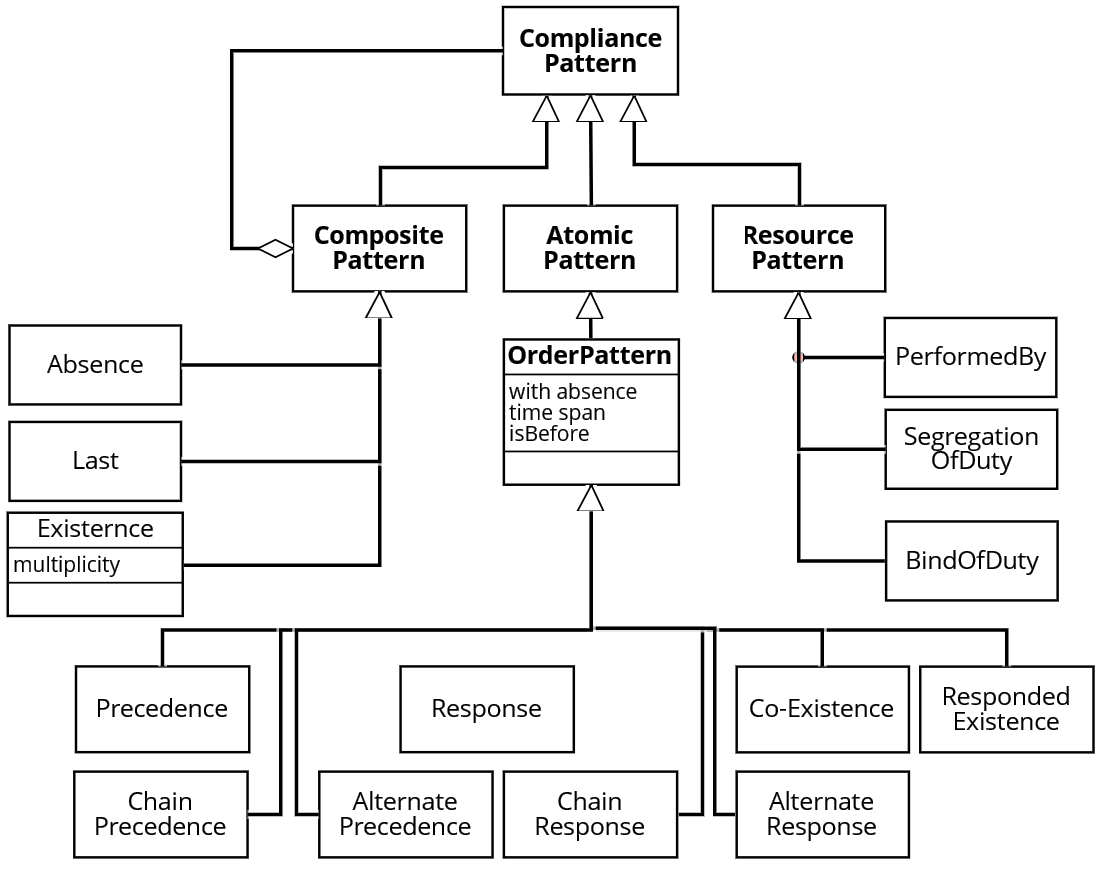}
\caption{Compliance Patterns}
\label{fig:compliance:pattern}    


\end{figure}

According to~\cite{elgammal2016formalizing}, the classification of patterns for business process compliance are \emph{Occurrence patterns}, \emph{Order patterns} and \emph{Timed patterns}, cf. Figure~\ref{fig:compliance:pattern}. These patterns are also common for modeling, discovering, or checking conformance of declarative process models~\cite{multiPerspectiveDECLARE16}.

\emph{Occurrence patterns} are concerned with activities having been executed (\emph{Existence}) or not (\emph{Absence}) within a process instance. \emph{Order patterns} are concerned with the execution order between pairs of activities, e.g., \emph{Precedes}, \emph{Response}. Besides the restriction on the execution order, the pattern can enforce further constraints. For example, it might define a time span (window) between the occurrence of the two activities and it might forbid other activities from occurring between the two events (see With Absence) in Figure~\ref{fig:compliance:pattern}. This latter restriction is also referred to as excluded activities.

The \emph{Response} pattern between two activities $A$ and $B$ requires that when $A$ occurs, $B$ should subsequently happen after $A$ in the same case before the process instance is terminated. Conversely, the \emph{Precedes} pattern implies that activity $B$ should occur only if $A$ has already occurred before for the same trace.

\emph{Alternate response} and \emph{Alternate precedence} restrict the \emph{Response} and \emph{Precedes} patterns by requiring activities to alternate without repetitions. \emph{Chain response} and \emph{Chain precedence} patterns specify even stronger ordering relations. These patterns require that the occurrences of $A$ and $B$ be close together. Table~\ref{tbl:LTL:formula} formalizes the compliance patterns using $LTL_f$ as common in literature. 

\begin{table}[ptb!]
\centering
    \footnotesize
    \caption{Formalization of conformance patterns using $LTL_{f}$ as common in literature}
    \vspace{0.3 cm}
\begin{tabular}{>{\columncolor[HTML]{EFEFEF}}l|l}
\hline
Pattern   & \cellcolor[HTML]{EFEFEF} \textbf{$LTL_{f}$ Formula}   \\ \hline
Absence & $ \neg \Diamond \phi$    
\\ \hline

Existence & $ \Diamond \phi$   
\\ \hline
Last & $\square (\phi \rightarrow \neg \bigcirc \neg \phi)$   
\\ \hline
Response & $\square (\phi \rightarrow \Diamond \psi)$ 
\\
Alter. Resp. & $\square (\phi \rightarrow \bigcirc(\neg \phi \UntilLTL \psi))$ 
\\ 
Chain Resp. & $\square (\phi \rightarrow \bigcirc(\psi))$ 
\\ \hline
Precedence & 
$(\neg \psi \UntilLTL \phi) \vee \square (\neg \psi)$ 
\\ 
Alter. Preced. & 

$ \square(\psi \rightarrow \bigcirc^{-1}(\neg \psi \UntilLTL \phi)))$  \\ 
Chain Preced. & 
$\square (\bigcirc(\psi) \rightarrow \phi)$ 
\\ \hline
Choice & 
$\Diamond \phi \vee \Diamond \psi$ 
\\ \hline
Resp. Exist. & 
$\Diamond \phi \rightarrow \Diamond \psi$ 
\\ \hline
Co-Existence & 
$\Diamond \phi \leftrightarrow \Diamond \psi$ 
\\ \hline
\end{tabular}
\label{tbl:LTL:formula}
\end{table}

\emph{Timed patterns} incorporate explicit time constraints into the compliance requirements in which an activity is expected to start/complete execution, e.g., \emph{within} the time interval in which an event or activity must take place or be finished or \emph{after} the time interval (\emph{isBefore} property in Figure~\ref{fig:compliance:pattern}). 

\begin{figure}[ptb!]
	\centering
		\includegraphics[width=\linewidth]{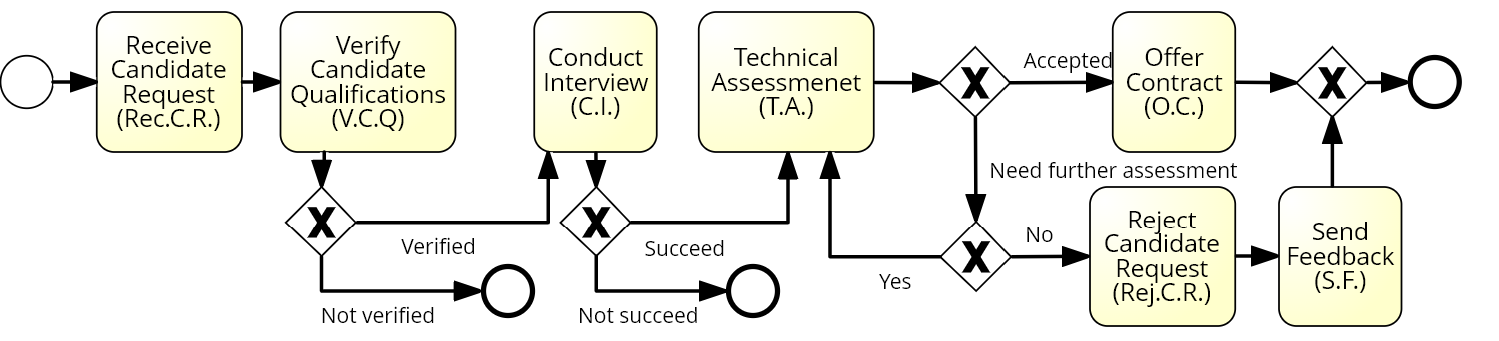}
		\caption{HR hiring process}
		\label{fig:running:example}    
\end{figure}

%
%
%

As discussed earlier, $LTL_f$ is incapable of capturing explicit timed patterns. On the other hand, MTL can capture these constraints. Following the \emph{With Absence}, \emph{Time Span}, and \emph{isBefore} properties of \emph{Ordered Patterns} in Figure~\ref{fig:compliance:pattern}, we can generalize, e.g., the response patterns to be further parameterized by $Ex$, the list of excluded activities, $I$, the time window (interval), and $\theta$, the comparison operator. Therefore, we can use the MTL formula  

\begin{equation}
\label{eq:example:MTL}
  \square_{\theta I}(a \rightarrow ((true \wedge \bigwedge_{e \in Ex} \neg e)  \UntilLTL_{\theta I} \Diamond b)  
\end{equation}


This formula ensures that whenever activity $a$ occurs, eventually activity $b$ occurs with respect to time interval $I$ according to $\theta$, meanwhile any activity $e\in Ex$ is not observed in-between. Note that we get the non-time restricted form of this pattern by setting $I=[0,\infty]$, $Ex = \phi$, and $\theta=\geq$. In this scenario, the formula is simplified to $\square (a \rightarrow true \UntilLTL b)$. Which can be simplified to $\square(a \rightarrow \Diamond b)$, similar to the formalization in Table~\ref{tbl:LTL:formula}.

When checking for compliance, analysts are interested in identifying process instances, i.e., traces that contain a violation, rather than those that are compliant. Therefore, it is common in the literature about compliance checking to use the term ``anti-pattern''~\cite{businessProcessModelAntiPattern2019} to refer to traces in which violations occur.


\subsection{A Running Example}
\label{sec:running:example}

An organization's Human Resources (HR) department needs to have a clear set of criteria when hiring an employee. There are several factors that can affect these criteria. For example, the organization should carefully evaluate the qualifications, experience, and other activities to ensure they meet the required standards and expectations. Moreover, it is important to take into account how well they align with the organization's culture and values.




\begin{table}[tb!]
    \centering
   \scriptsize
    \caption{A sample event log for hiring process with additional attributes}
    \label{tbl:hiringeventlog}
    \begin{tabular}{c|c|c|o|o|x}
        \hline
    	\textbf{Event} & \textbf{C.ID} & \textbf{Activity} & \textbf{StartTime}& \textbf{CompleteTime}  & \textbf{Position} \\
    	\hline 
            1 & 1 & \textbf{Start} & 10-03-2021 20:20 & 10-03-2021 20:20 & 0 \\
            2 & 1 & Rec.C.R. & 10-03-2021 20:26 & 13-03-2021 15:54 & 1 \\
	    3 & 1 & V.C.Q & 14-03-2021 09:24 & 14-03-2021 15:50 & 2 \\
      4 & 2 & \textbf{Start} & 09-03-2021 09:05 & 09-03-2021 09:05 & 0 \\
	    5 & 2 & Rec.C.R. & 09-03-2021 09:11 & 10-03-2021 13:53 & 1 \\
	    6 & 1 & C.I. & 16-03-2021 13:00 & 16-03-2021 15:11 & 3 \\
	 7 & 3 & \textbf{Start} & 31-03-2021 15:25 & 31-03-2021 15:25 & 0 \\   
     8 & 3 & Rec.C.R. & 31-03-2021 15:30 & 02-04-2021 13:11 & 1 \\
	    9 & 1 & T.A. & 18-03-2021 10:30 & 18-03-2021 12:30 & 4 \\
            10 & 2 & V.C.Q. & 15-03-2021 12:48 & 16-03-2021 17:08 & 2 \\
	    11 & 2 & T.A. & 19-03-2021 12:22 & 19-03-2021 14:22 & 3 \\
	    12 & 3 & V.C.Q. & 05-04-2021 13:34 & 07-04-2021 10:24 & 2 \\
	    13 & 2 & C.I. & 22-03-2021 14:20 & 22-03-2021 16:36 & 4 \\
            14 & 2 & O.C. & 30-03-2021 12:58 & 30-03-2021 14:56 & 5 \\
            15 & 2 & \textbf{End} & 30-03-2021 15:00 & 30-03-2021 15:00 & 5 \\
	    16 & 3 & C.I. & 09-04-2021 11:48 & 09-04-2021 13:19 & 3 \\
            17 & 4 & \textbf{Start} & 03-03-2021 14:10 & 03-03-2021 14:10 & 0 \\
            18 & 4 & Rec.C.R. & 03-03-2021 14:17 & 06-03-2021 10:39 & 1 \\
            19 & 1 & O.C. & 20-03-2021 13:33 & 20-03-2021 15:50 & 5 \\
            20 & 1 & \textbf{End} & 20-03-2021 15:55 & 20-03-2021 15:55 & 5 \\
            21 & 4 & C.I. & 10-03-2021 15:00 & 10-03-2021 16:41 & 2 \\
            22 & 3 & T.A. & 12-04-2021 12:39 & 12-04-2021 14:40 & 4 \\
            23 & 4 & V.C.Q. & 12-04-2021 14:05 & 14-04-2021 18:40 & 3 \\ 
            24 & 4 & T.A. & 18-04-2021 09:30 & 18-04-2021 11:30 & 4 \\
            25 & 3 & T.A. & 16-04-2021 10:32 & 16-04-2021 12:30 & 5 \\
            26 & 4 & O.C. & 20-04-2021 12:05 & 20-04-2021 12:19 & 5 \\
            27 & 4 & \textbf{End} & 20-04-2021 12:25 & 20-04-2021 12:25 & 5 \\
            28 & 3 & Rej.C.Q. & 18-04-2021 13:05 & 18-04-2021 13:15 & 6 \\ 
	      29 & 3 & S.F. & 19-04-2021 15:31 & 19-04-2021 15:55 & 7 \\
       30 & 3 & \textbf{End} & 19-04-2021 16:05 & 19-04-2021 16:05 & 7 \\
    \tikzmark{a1} \vdots & \vdots & \vdots\tikzmark{a2} & \tikzmark{b1}\vdots& \vdots\hspace{2mm} \tikzmark{b2} & \tikzmark{c1}\hspace{2mm}\vdots\hspace{2mm} \tikzmark{c2} \\
    	\hline 
	\end{tabular}
	\begin{tikzpicture}[overlay, remember picture]
        \draw [decorate,decoration={brace,amplitude=10pt,mirror,raise=4pt}] (a1.west) --node[below=14pt]{Minimum details} (a2.east);
        \draw [decorate,decoration={brace,amplitude=10pt,mirror,raise=4pt}] (b1.west) --node[below=14pt]{Optional details} (b2.east);
        \draw [decorate,decoration={brace,amplitude=10pt,mirror,raise=4pt}] (c1.west) --node[below=14pt]{Added detail} (c2.east);
    \end{tikzpicture}
    
  \vspace{7mm}
      
\end{table}

Figure~\ref{fig:running:example} depicts the typical activities involved in hiring an employee and the order in which they should be executed as a BPMN process model. Table~\ref{tbl:hiringeventlog} shows a sample event log for such a hiring process. It has an event identification number ($Event$), here presented in serial numbers, a case identifier ($C.ID$) that presents a unique number per employment candidate, and $Activity$ presents the action performed in the hiring process. ``StartTime'' and ``CompleteTime'' columns are considered optional details. Finally, the ``Position'' column specifies the activity's position per case. 

Table~\ref{tbl:complianceRequirement} shows an excerpt of the compliance requirements enforced upon the process and that needs to be checked against the log (Table~\ref{tbl:hiringeventlog}). Table~\ref{tbl:complianceRequirement} shows that the Linear Temporal Logic ($LTL_{f}$) formula cannot express $Req.2~\&~Req.4$ due to temporal constraints.



%

\begin{table}[ptb!]
    \centering
    \caption{An Excerpt of the Compliance Requirements}
    \label{tbl:complianceRequirement}
    \footnotesize
\begin{tabular}{p{0.7cm} p{6cm} p{4cm}}
\hline
\rowcolor[HTML]{C0C0C0} 
\textbf{ID} & \textbf{Compliance Requirements} & \textbf{$LTL_{f}$ formula}\\ \hline
Req.1       &   Provide constructive feedback to candidates who were interviewed but not selected for the position &  $\square (Rej.C.Q \rightarrow \Diamond S.F.)$                            \\ \hline
Req.2       & The HR must verify the candidate's qualifications within 2 days of receiving the application.      &  $\textbf{\textemdash}$                          \\ \hline
Req.3       &  The candidate must successfully complete the interview before undergoing the technical assessment.                           & $(\neg T.A.~\UntilLTL~C.I.) \vee \square (\neg T.A.)$      \\ \hline
Req.4       &  For any candidate who successfully passes the technical assessment will receive the job offer 3 days later.                  & $\textbf{\textemdash}$        \\ \hline
Req.5       &  Upon the verification of candidate qualifications, the candidate is eligible to proceed to an interview before entering the technical assessment phase.       &   $\square (V.C.Q. \rightarrow \bigcirc(\Diamond C.I.~\UntilLTL~T.A.))$              \\ \hline
\end{tabular}
\end{table}

\section{Related Work}\label{sec:related:work}

In recent years, an increasing number of researchers have focused on declarative conformance checking.  In~\cite{montali2010declarative}, the authors focus on ensuring compliance with business rules by translating constraints into Linear Temporal Logic (LTL) and subsequently assessing them through the use of automata. 

In~\cite{maggi2011}, the authors introduce an initial method for mining declarative process models using Declare constraints. This method has improved in~\cite{maggi2012}, adopting a two-phase strategy. Initially, the first phase employs an apriori algorithm to detect frequent sets of interconnected activities. From these sets, a list of potential constraints is derived. Subsequently, in the second phase, these constraints are evaluated by replaying the log against specific automata, each designed to accept traces adhering to a specific constraint. Constraints that satisfy a predefined threshold, determined by the percentage of traces, are then identified and considered as discovered constraints. 

In another research, the authors introduce a method designed to assess the alignment of a log with a Declare model~\cite{burattin2012techniques}. Their algorithm determines, for each trace, whether a Declare constraint is violated or satisfied. These methods focus on the control-flow perspective and do not consider the data nor time perspectives. In~\cite{connectingDBPM19}, the authors present a toolset to extract process-related data from operational logs (e.g., relational databases) using a metamodel. It can represent complex SQL queries involving nesting and joins. 

Relational databases have also been used in declarative process mining for compliance verification. SQLMiner, as introduced in~\cite{declarativeMiningSQL16}, simplifies this process by efficiently using SQL queries to extract compliance patterns from event data. Sch{\"o}nig et al.~\cite{schonig2016discovery} present a framework developed to discover MP-Declare models using SQL queries. This framework can analyze processes from multiple viewpoints, including control flow, data attributes, organizational aspects, and time perspectives. 
 
Compliance violations, i.e., anti-patterns, can be checked by \emph{Match\_Recognize} (MR), the ANSI SQL operator. MR verifies patterns as regular expressions, where the tuples of a table are the symbols of the string to search for matches within. MR runs linearly through the number of tuples in the table. In our case, the tuples are the events in the log. In practice, the operational time can be enhanced by parallelizing the processing, e.g., partitioning the tuples by the case identifier. Still, this does not change the linearity of the match concerning the number of tuples in the table. A recent work speeds up MR by using indexes in relational databases for strict contiguity patterns, i.e., patterns where events are in strict sequence~\cite{indexAcceleratedPatternMatching21}. However, Order compliance patterns frequently refer to eventuality rather than strict order, limiting the use of indexes to accelerate the matching process.

Leveraging multi-perspective declarative process models and advanced conformance checking techniques add competitive advantages to organizations. Organizations can gain insights into process compliance, identify bottlenecks, improve process efficiency, and ensure adherence to defined business rules and constraints across different dimensions of the process. In ~\cite{multiPerspectiveDECLARE16}, the author design an approach to conduct conformance checking with declarative process models based on log replay. They developed a model interpreter to extract linear temporal logic $LTL$ constraints
from the input Declare model. Subsequently, the approach checks each trace to determine whether these constraints are violated or not. They support $Declare~Analyzer$ methods to detect constraints.


Storing and querying event logs using graph databases has also been investigated. Esser et al.~\cite{multiDimlEventGrahDB21} provide a rich, multidimensional model for event data using labelled property graphs realized on top of Neo4J. The authors show how their model supports several classes of queries on event data. All compliance patterns can be represented as queries against their model. To conclude, this area is still open for improvement due to the complexity of constraints in real-life processes and the need for risk management and control to minimize violations.

\section{Formalization of Compliance Patterns using $MTL_{f}$}
\label{sec:formalization}

We have seen in Section~\ref{sec:running:example} that $LTL_f$ falls short of capturing timed patterns, whereas $MTL_f$ easily captures the timed and untimed compliance patterns. In this section, we detail our first contribution where we provide a minimal set of $MTL_f$ formulas that we argue it covers all of the known occurrence, order, or timed patterns. The purpose of having this minimal set is to modularize the mapping from the temporal logic formalism to the different query languages suitable to access event logs stored in the respective data model (Section~\ref{sec:proposed:approach}).

In general, we can extend most of the patterns discussed in Section~\ref{sub:sec:compliance:patterns} by defining the time interval $I$ in which the constraint is required to hold. Therefore, using $MTL_f$ instead of $LTL_f$. Moreover, we emphasize the use of the set of excluded activities as it will play a crucial role in simplifying the formulas and reducing their numbers while capturing the patterns. We will use the template $Pattern(condition, target, Excluded, Interval)$ to parameterize the patterns. For instance, the \emph{response} pattern is defined as $Response(\phi,\psi, \Gamma, I)$. The \emph{absence} pattern is defined as $Absence(\bot, \bot, \{\gamma\},I)$. We will elaborate more on this in the following subsections.



\subsection{Generalized $MTL_f$ Formulas} \label{sub:sec:formal:compact}


Formula~\ref{General:MTL:formula} is a $MTL_f$ template formula that covers future-looking compliance patterns.

\begin{equation}
\label{General:MTL:formula}
\square ((S ~\vee \phi) \rightarrow \bigcirc( (True ~\wedge \bigwedge_{\gamma \in \Gamma} \neg~\gamma) \UntilLTL_{I} (\psi ~\vee E)))
\end{equation}

The template in Formula~\ref{General:MTL:formula} is parameterized by activities  $\phi,~\psi,~$and$~\Gamma$; where $\phi\in D_a,~\psi\in D_a$, and $\Gamma \subset D_a$, c.f. Definition~\ref{def:event}, and the time interval $I$. $\phi$ is the condition for the implication, $psi$ is the consequent, whereas $\Gamma$ is the set of excluded activities. $\Gamma$ and $I$ are optional. If they are not specified, the resulting formula reduces to an $LTL_f$ formula. There are two special activities $S,E \notin D_a$ that are used to indicate the start $S$, and the end $E$ of a trace (case). This is a common preprocessing approach for event logs. Formula~\ref{General:MTL:formula} also generalizes Formula~\ref{eq:example:MTL}, by adding the special events $S$ and $E$. 

Formula~\ref{General:MTL:formula} guarantees that when the activation activity $\phi$ is executed, the target activity $\psi$ will hold at some future point within a specific time interval $I$. In between, no $\gamma \in \Gamma$ shall be executed. Formula~\ref{General:MTL:formula} uses future operators, i.e. $\bigcirc$ and $\UntilLTL$. Although such operators do not improve the expressive power of temporal logic~\cite{LTLFairness}, they still make temporal logic formulas more succinct. Therefore, Formula~\ref{General:MTL:formula:past} is the past-looking template that is used to cover the precedence pattern variants.  In Formula~\ref{General:MTL:formula:past}, $\phi$ and $\psi$ swap their roles for being the condition and the consequent.

\begin{equation}
\label{General:MTL:formula:past}
\square ((E ~\vee \psi) \rightarrow \bigcirc^{-1}( (True ~\wedge \bigwedge_{\gamma \in \Gamma} \neg~\gamma) \UntilLTL^{-1}_{I} (\phi ~\vee S)))
\end{equation}

Table~\ref{tbl:MTL:formula} shows the instantiation of Formulas~\ref{General:MTL:formula} and~\ref{General:MTL:formula:past} to model the different compliance patterns from Table~\ref{tbl:LTL:formula} in $MTL_{f}$. 
For instance, to model the absence pattern $Absence(\bot, \bot,\{\gamma\}, I)$, Formula~\ref{General:MTL:formula} is instantiated by defining the excluded events $\Gamma =\{\gamma\}$, dropping $\phi$, $\psi$, and $True$ to get $\square ((S) \rightarrow \bigcirc( (\neg \gamma) \UntilLTL_{I} (E)))$. The start event $S$ holds in the first state, and from the next state, no $\gamma$ shall be observed until the end activity $E$ is reached.

\break

\begin{table}[pt!]
\centering
\small
    \caption{Generalization of $LTL_{f}$-based Conformance Patterns in $MTL_{f}$ Formula}
    \label{tbl:MTL:formula}
\begin{tabular}{>{\columncolor[HTML]{EFEFEF}}l|l}
\hline
Pattern     & \cellcolor[HTML]{EFEFEF} General $MTL_{f}$ Formula    \\ \hline
Absence($\bot,\bot,\{\gamma\}, I$) &$\square (S \rightarrow \bigcirc( \neg \gamma \UntilLTL_{I} E))$  
\\ \hline

Existence($\bot,\psi, \{\},I$) & $\square (S \rightarrow \bigcirc(T \UntilLTL_{I} \psi))$  
\\ \hline
Last($\phi,\bot,D_a,I$)  & $\square (\phi \rightarrow \bigcirc(\bigwedge_{\gamma \in D_a \setminus \{\phi \}} \neg \gamma \UntilLTL_{I} E)))$  
\\ \hline
Response($\phi,\psi,\Gamma,I$) & $\square (\phi \rightarrow \bigcirc ((True \wedge \bigwedge_{\gamma \in \Gamma}  \neg \gamma) \UntilLTL_{I} \psi))$
\\
Alter. Resp.($\phi,\psi,\Gamma,I$)  & Response($\phi,\psi,\Gamma \cup \{\phi\},I$)

\\ 
Chain Resp.($\phi,\psi,\{\},I$)  & Response($\phi,\psi,D_a \setminus \{\psi\},I$)
\\ \hline
Precedence($\psi$,$\phi$,$\Gamma$,$I$) & $\square (\psi \rightarrow \bigcirc^{-1}((T \wedge \bigwedge_{\gamma \in \Gamma} \neg \gamma) \UntilLTL^{-1}_{I}\phi))$
\\ 
Alter. Preced.($\psi,\phi,\Gamma,I$) & Precedence($\psi,\phi,\Gamma \cup \{\psi\},I$)

\\ 
Chain Preced.($\psi,\phi,\{\},I$)  & Precedence($\psi,\phi,D_a \setminus \{\phi\},I$)

\\ \hline
Choice($\phi,\psi,\{\},I$) & $Existence(\bot, \phi,\{\},I) \vee Existence(\bot, \psi,\{\},I)$
\\ \hline
Resp. Exist.($\phi,\psi,\{\},I$) & $Existence(\bot, \phi,\{\},I) \rightarrow Existence(\bot, \psi,\{\}, I)$\\ \hline
Co-Existence($\phi,\psi,\{\},I$) & $Existence(\bot, \phi,\{\},I) \leftrightarrow Existence(\bot, \psi,\{\}, I)$
\\ \hline
\end{tabular}

\end{table}

The existence patterns $Existence(\bot, \psi, \{\}, I)$ can be modeled as $\psi$ is the consequent for observing $S$. To model the last pattern, $Last(\phi, \bot, \Gamma, I)$, we set $\Gamma = D_a$. We drop $S$ from the formula as we want to check for the occurrence of $\phi$ as a condition. Once $\phi$ holds in some state, in all future states, no activity from $D_a$ shall be observed. This effectively means that in the next state, only $E$ should be observed for the trace to be compliant.  

The $Response(\phi,\psi, \Gamma, I)$ is straightforward. The alternating response is modeled as $AltResp(\phi,\psi, \Gamma, I)$ using the response pattern by adding $\phi$ to $\Gamma$, the set of excluded events. $ChainResp$ uses the response pattern formula, where $\Gamma = D_a \setminus \{\psi\}$. This effectively means that once $\phi$ is observed in some state, no other activity than $\psi$ can be observed in the immediate next state.

Likewise, the precedence pattern variants can be built on the basic precedence pattern. The last three patterns in Table~\ref{tbl:MTL:formula} are built on top of the \emph{existence} pattern using different logical operators.






\subsection{Reflection on the Running Example}


\begin{table}[htb!]
    \centering
    \caption{Formalization of the Compliance Requirements using Formula~\ref{General:MTL:formula}}\label{tbl:compliance:req:MTL}
\small
\begin{tabular}{c l}
\hline

\rowcolor[HTML]{C0C0C0} 
\textbf{ID} & \textbf{$MTL_{f}$ Formula} \\ \hline
Req.1       &  $\square (\textbf{C.I.} \rightarrow \bigcirc (True ~ \UntilLTL_{[0,\infty]} ~ \textbf{S.F.}))$ 
      \\ \hline
Req.2       & $\square (\textbf{V.C.Q.} \rightarrow \bigcirc^{-1}(True ~ \UntilLTL^{-1}_{[0,48]} ~ \textbf{Rec.C.R.}))$
                             \\ \hline
Req.3       &  $\square (\textbf{T.A.} \rightarrow \bigcirc^{-1} (True ~ \wedge \neg \textbf{V.C.Q.} ~ \UntilLTL^{-1}_{[0,\infty]} ~ \textbf{C.I.}))$
                              \\ \hline
Req.4       &  $\square (\textbf{T.A.} \rightarrow \bigcirc (True ~ \UntilLTL_{[72,\infty]} ~ \textbf{O.C.}))$
\\ \hline
Req.5       &  $\square (\textbf{V.C.Q.} \rightarrow \bigcirc (True ~ \wedge \neg \textbf{T.A.} ~ \UntilLTL_{[0,\infty]} ~ \textbf{C.I.}))$
                    \\ \hline
\end{tabular}

\end{table}

Table~\ref{tbl:compliance:req:MTL} shows how the compliance requirements from Section~\ref{sec:running:example} are modeled following the pattern templates in Table~\ref{tbl:MTL:formula}.
Table~\ref{tbl:hiringeventlog} can help to illustrate the above constraints. For instance, by examining the \emph{Precedence} constraint, we can determine that case $2$ did not conform to the time restriction provided in Req.2. Similarly, in Req.5 (\emph{Response Constraint}), we must verify that activity $C.I.$ follows activity $V.C.Q.$ with no instances of activity $T.A.$ in between. This requirement is not fulfilled in case $2$. Additionally, in Req.3, case $4$ does not satisfy the \emph{Alternate Precedence} constraint due to the occurrence of the activity \emph{V.C.Q.}, which violates the constraint. 

Section~\ref{sec:proposed:approach} examines the compatibility of compliance patterns across various data models.



\section{Mapping $MTL_{f}$-based Compliance Patterns To the Different Data Models}\label{sec:proposed:approach}

In this section, we aim to present the conversion of Metric Temporal Logic $(MTL_{f})$ compliance patterns into different data models, namely the Relational Data Model and Graph Data Model. This translation process entails representing the temporal constraints using the query language of each data model.

In the mapping, however, we care about mapping the anti-pattern. That is, the negation of the respective $MTL_f$ pattern~\cite{DBLP:conf/sac/AwadBESAS15,Antipatterns}. That is because, we care about identifying the violation(s).  Moreover, due to space limitations, we cover the anti patterns derived from the Formula~\ref{General:MTL:formula}. The anti-patterns derived from Formula~\ref{General:MTL:formula:past} can be derived in a similar way but using the converse comparison operators.

\subsection{Relational Data Model}

The log representation in Table~\ref{tbl:hiringeventlog} maps directly to a table in an RDBMS. The columns for case identifier, event identifier, activity, etc. can be stored directly in the relational database. The SQL Miner~\cite{declarativeMiningSQL16} technique uses SQL operators like joins and nested queries to identify compliance violations. 

The SQL standard defines the \texttt{Match\_Recognize} (MR) pattern matching operator. The patterns finds matches a table's records the same way a regular expression finds matches over a string. However, not all RDBMSs implement the MR operator. Nevertheless, we consider mapping $MTL_f$ to MR over a relational table storing the event log.

Another method uses the analytical Match Recognize (MR) operator. Section~\ref{sec:SQL:Miner} and Section~\ref{sec:Match:Recognize} details the queries for each approach, respectively.

\subsubsection{Mapping using SQL Miner}\label{sec:SQL:Miner}

The following queries show different variations of the response anti-pattern query using SQL Miner, either within time interval $I$ (listing~\ref{query7}), after time interval $I$ (listing~\ref{query8}), or without time interval (listing~\ref{query9}), respectively. For instance, the query in Listing~\ref{query7} shows \emph{Response} anti-pattern within specified time frame $I$. It reports violations that occurr either due to the time frame was not met, in Line $4$, or the absence of the target activity $\psi$ as in Line 15. It also detects cases where excluded activities $\Gamma$  are observed between activities $\phi$ and $\psi$, as in Line $5$. All other listings are working in the same way with different representation.

\begin{lstlisting}[
language=SQL,
deletekeywords={IDENTITY},
deletekeywords={[2]INT},
morekeywords={clustered},
label={query7},
caption={Response anti-pattern query within time interval (I) using SQL Miner},escapeinside={(*@}{@*)}]
Select CID 
From log as l, log as l1 
Where l.CID = l1.CID and l.activity = (*@$\phi$@*) and l1.activity = (*@$\psi$@*)
and l1.position > l.position and (l1.start-l.complete >= I
or EXISTS (select CID FROM log as l2
      WHERE l2.CID = l.CID
      AND l2.position > l.position
      AND l2.position < l1.position
      AND l2.activity IN ((*@$\Gamma$@*)) 
      )
Union
Select CID 
From log as l
Where l.activity = (*@$\phi$@*) 
and l.CID not in (Select l1.CID From log l1 Where l1.activity = (*@$\psi$@*))
\end{lstlisting} 

\begin{lstlisting}[
language=SQL,
deletekeywords={IDENTITY},
deletekeywords={[2]INT},
morekeywords={clustered},
label={query8},
caption={Response anti-pattern query after time interval (I) using SQL Miner},escapeinside={(*@}{@*)}]
Select CID 
From log as l, log as l1 
Where l.CID = l1.CID and l.activity = (*@$\phi$@*) and l1.activity = (*@$\psi$@*)
and l1.position > l.position and (l1.start-l.complete <= I
or EXISTS (select CID FROM log as l2
      WHERE l2.CID = l.CID
      AND l2.position > l.position
      AND l2.position < l1.position
      AND l2.activity IN ((*@$\Gamma$@*)) 
      ) 
    )
Union
Select CID 
From log as l
Where l.activity = (*@$\phi$@*) 
and l.CID not in (Select l1.CID From log l1 Where l1.activity = (*@$\psi$@*))
\end{lstlisting} 

\begin{lstlisting}[
language=SQL,
deletekeywords={IDENTITY},
deletekeywords={[2]INT},
morekeywords={clustered},
label={query9},
caption={Response anti-pattern query using SQL Miner},escapeinside={(*@}{@*)}]
Select CID 
From log as l, log as l1 
Where l.CID = l1.CID and l.activity = (*@$\phi$@*) and l1.activity = (*@$\psi$@*)
and l1.position < l.position
and EXISTS (select CID FROM log as l2
      WHERE l2.CID = l.CID
      AND l2.position > l.position
      AND l2.position < l1.position
      AND l2.activity IN ((*@$\Gamma$@*)) 
      )
Union
Select CID 
From log as l
Where l.activity = (*@$\phi$@*) 
and l.CID not in (Select l1.CID From log l1 Where l1.activity = (*@$\psi$@*))
\end{lstlisting} 

\subsubsection{Mapping using Match Recognize}
\label{sec:Match:Recognize}

The following queries show different variations of the response anti-pattern query using Match Recognize, either within time interval $I$ (listing~\ref{query10}), after time interval $I$ (listing~\ref{query12}), or without time interval (listing~\ref{query11}), respectively. Let's explain how the query in listing \ref{query10} detects violations. Basically, it defines the pattern, in Line $7$, where $A$ defined as an activation activity $\phi$, in Line $9$, and $B$ as a target activity $\psi$, in Line $11$ or $E$ as an End activity, in Line $25$. Additionally, the set $S^*$ contains all excluded activities, as indicated in Line $10$. The violations occurs under two conditions: first, when the time frame is not met, in Line $12$, and second the absence of the target activity $\psi$, as in Line $24$. The remaining listing queries work in the same manner  with different representations.

\break

\begin{lstlisting}[
language=SQL,
deletekeywords={IDENTITY},
deletekeywords={[2]INT},
morekeywords={clustered},
label={query10},
caption={Response anti-pattern query within time interval (I) using Match Recognize},escapeinside={(*@}{@*)}]
select distinct CID
from log MATCH_RECOGNIZE(
     PARTITION BY CID
     ORDER BY Time_stamp
     ONE ROW PER MATCH
     AFTER MATCH SKIP TO NEXT ROW
     PATTERN (A S* B)
     DEFINE
              A AS Activity_ID = (*@$\phi$@*),
              S AS S.Activity_ID <> (*@$\phi$@*) AND S.Activity_ID <> (*@$\psi$@*) AND S.Activity_ID <> (*@$\Gamma$@*),
              B AS ((B.Activity_ID = (*@$\psi$@*) 
              AND  (B.Time_stamp - A.time_stamp) > interval I)) 
     )
union
select distinct CID
from log MATCH_RECOGNIZE(
     PARTITION BY CID
     ORDER BY Time_stamp
     ONE ROW PER MATCH
     AFTER MATCH SKIP TO NEXT ROW
     PATTERN (A S* E )
     DEFINE
              A AS Activity_ID = (*@$\phi$@*),
              S AS S.Activity_ID <> (*@$\phi$@*) AND S.Activity_ID <> (*@$\psi$@*),
              E AS E.Activity_ID = 'END'
     );     
\end{lstlisting} 


\begin{lstlisting}[
language=SQL,
deletekeywords={IDENTITY},
deletekeywords={[2]INT},
morekeywords={clustered},
label={query12},
caption={Response anti-pattern query after time interval (I) using Match Recognize},escapeinside={(*@}{@*)}]
select distinct CID
from log MATCH_RECOGNIZE(
     PARTITION BY CID
     ORDER BY Time_stamp
     ONE ROW PER MATCH
     AFTER MATCH SKIP TO NEXT ROW
     PATTERN (A S* B)
     DEFINE
              A AS Activity_ID = (*@$\phi$@*),
              S AS S.Activity_ID <> (*@$\phi$@*) AND S.Activity_ID <> (*@$\psi$@*) AND S.Activity_ID <> (*@$\Gamma$@*),
              B AS ((B.Activity_ID = (*@$\psi$@*) 
              AND  (B.Time_stamp - A.time_stamp) < interval I)) 
     )
union
select distinct CID
from log MATCH_RECOGNIZE(
     PARTITION BY CID
     ORDER BY Time_stamp
     ONE ROW PER MATCH
     AFTER MATCH SKIP TO NEXT ROW
     PATTERN (A S* E )
     DEFINE
              A AS Activity_ID = (*@$\phi$@*),
              S AS S.Activity_ID <> (*@$\phi$@*) AND S.Activity_ID <> (*@$\psi$@*),
              E AS E.Activity_ID = 'END'
     );     
\end{lstlisting}

\break

\begin{lstlisting}[
language=SQL,
deletekeywords={IDENTITY},
deletekeywords={[2]INT},
morekeywords={clustered},
label={query11},
caption={Response anti-pattern query using Match Recognize},escapeinside={(*@}{@*)}]
select distinct CID
from log MATCH_RECOGNIZE(
	PARTITION BY CID
	ORDER BY Time_stamp
	ONE ROW PER MATCH
	AFTER MATCH SKIP TO NEXT ROW
	PATTERN (A S* B)
	DEFINE
		A AS Activity_ID = (*@$\phi$@*),
		S AS S.Activity_ID <> (*@$\phi$@*) AND S.Activity_ID <> (*@$\psi$@*) AND S.Activity_ID = (*@$\Gamma$@*),
		B AS B.Activity_ID = (*@$\psi$@*) 
	)
union
select distinct CID
from log MATCH_RECOGNIZE(
	PARTITION BY CID
	ORDER BY Time_stamp
	ONE ROW PER MATCH
	AFTER MATCH SKIP TO NEXT ROW
	PATTERN (A S* E )
	DEFINE
		A AS Activity_ID = (*@$\phi$@*),
		S AS S.Activity_ID <> (*@$\phi$@*) AND S.Activity_ID <> (*@$\psi$@*),
		E AS E.Activity_ID = 'END'
	);     
\end{lstlisting}

\subsection{Graph Data Model}

Graph representation model of event logs is a potential way for analyzing event logs ~\cite{processAtlas18}, particularly for compliance checking~\cite{multiDimlEventGrahDB21}, due to the richness of the graph data model, mature database engines supporting it, e.g., Neo4J~\footnote{\scriptsize Neo4J website \url{https://neo4j.com/}}, and the declarative query languages embraced by such engines, e.g., Cypher~\footnote{\scriptsize Cypher for Neo4J is like SQL for relational databases.}. In this manner, compliance checking can be mapped to queries against the encoded log to detect violations. For instance, a \emph{precedence} pattern of activity A over activity B within a specific time frame is expressed by identifying nodes corresponding to activities A and B. The verification process is checking the relationship between these nodes via directed edges to ensure that it meets the time constraints. There are two different encoding methods: \emph{Multi-Dimensional method}~\cite{multiPerspectiveDECLARE16} and \emph{Unique Activities method} \cite{Antipatterns}, outlined in Sections \ref{sec:baseline} and \ref{sec:unique}, respectively.

\subsubsection{Mapping to Multi-Dimensional Encoding} \label{sec:baseline}
\begin{figure}[hpbt!]
\centering
\includegraphics[width=0.85\linewidth]{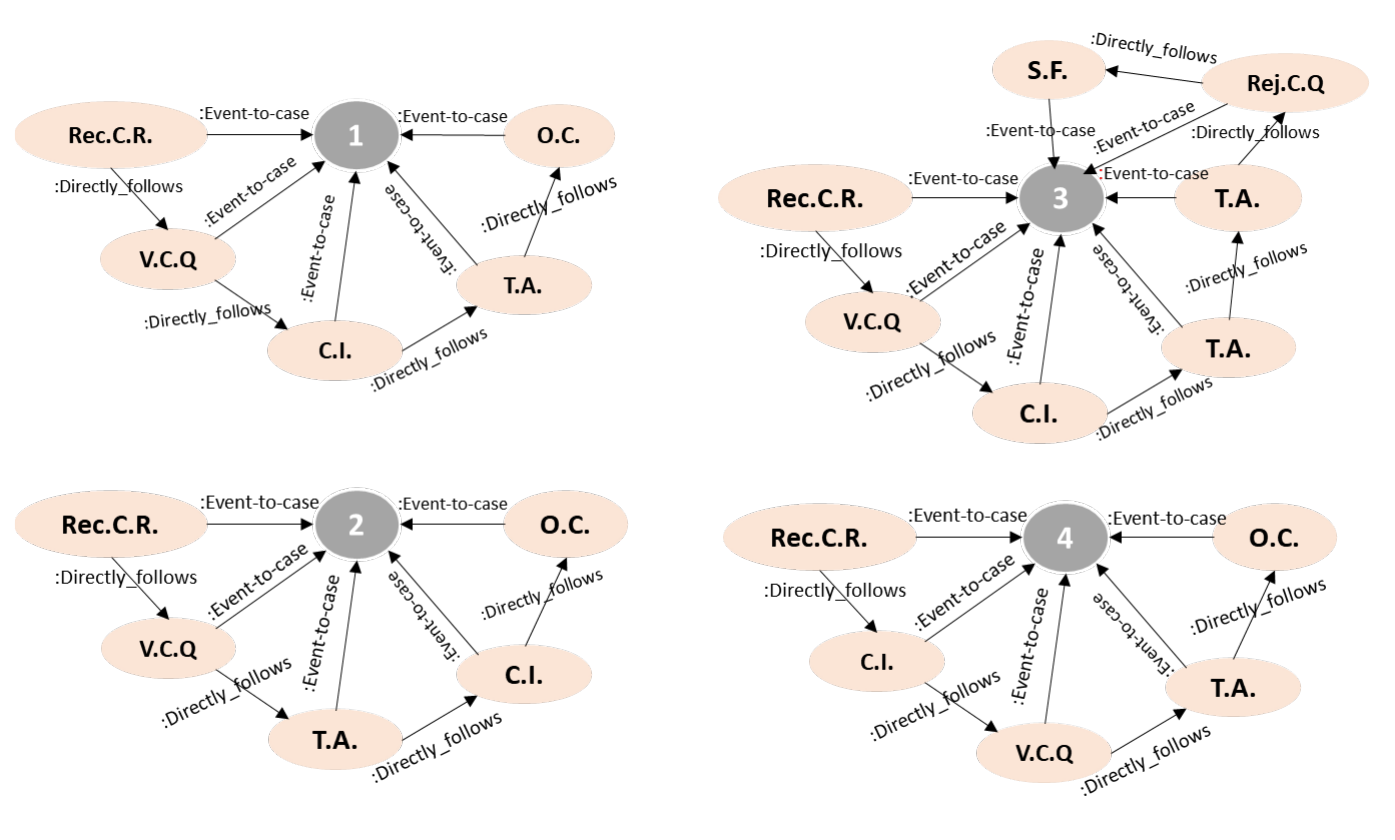}
\caption{Multi-Dimensional Encoding of~Table~\ref{tbl:hiringeventlog}}
\label{fig:multi:activities:encodings}    
\end{figure}

This approach was one of the graph encoding methods. The disadvantage of this approach, especially when employing linear scan, is leading to performance issues. As the size or complexity of the graph increases, the linear scan method might lead to inefficiencies, significantly slowing down the query or analysis processes. 
Figure~\ref{fig:multi:activities:encodings} depicts the graph produced by encoding the log excerpt in Table~\ref{tbl:hiringeventlog} using the Multi-Dimensional method. 

Listings \{\ref{query4},\ref{query5},\ref{query6}\} demonstrate the usage of \emph{Cypher} queries to define compliance patterns under time constraints. 
The query in Listing \ref{query4} expresses \emph{Response} anti-pattern within specific time frame. First, the query identifies the beginning of each trace \texttt{(start:Event\{activity:$\phi$\})}. Then the sequence of nodes constituting a path from each node of activity End to the start activity $\phi$, in the same case, is constructed. The path is constructed by traversing the transitive closure of the :Directly follows relation; it is represented as \texttt{path=(e1:Event\{activity:End\}<-[:Directly\_follows*]- (start))}. If the path does not meet time frame, Line 5, or does not include any node whose activity property refers to $\psi$, line 7, a violation exists, and this case is reported. Furthermore, if the path includes excluded activities $\Gamma$, Line 6, the violation exists. The same applies for the other two listings with different representation of time interval inclusion or exclusion.


\begin{lstlisting}[
language=SQL,
deletekeywords={IDENTITY},
deletekeywords={[2]INT},
morekeywords={clustered},
label={query4},
caption={Response anti-pattern query within time interval (I) using the Mutli-Dimensional encoding},escapeinside={(*@}{@*)}]
Match (c:Case) <-[:Event_to_case]- (start:Event{activity:(*@$\phi$@*)})
Match (e1:Event{activity:(*@$\psi$@*)}<-[:Directly_follows*]-(start)
Match path = (e1:Event{activity:End}<-[:Directly_follows*]-(start)) 
	where exists (n in nodes(path) where n.activity=(*@$\psi$@*)
	and n.startTime- start.completeTime >= I) 
	or (start)-[:Directly_follows*]->((*@$\Gamma$@*))-[:Directly_follows*]->(e1) 
	or none (n in nodes(path) where n.activity=(*@$\psi$@*))
return c.ID
\end{lstlisting}  

\begin{lstlisting}[
language=SQL,
deletekeywords={IDENTITY},
deletekeywords={[2]INT},
morekeywords={clustered},
label={query5},
caption={Response anti-pattern query after time interval (I) using the Mutli-Dimensional encoding},escapeinside={(*@}{@*)}]
Match (c:Case) <-[:Event_to_case]- (start:Event{activity:(*@$\phi$@*)})
Match (e1:Event{activity:(*@$\psi$@*)}<-[:Directly_follows*]-(start)
Match path = (e1:Event{activity:End}<-[:Directly_follows*]-(start)) 
	where exists (n in nodes(path) where n.activity=(*@$\psi$@*)
	and n.startTime- start.completeTime <= I) 
	or (start)-[:Directly_follows*]->((*@$\Gamma$@*))-[:Directly_follows*]->(e1) 
	or none (n in nodes(path) where n.activity=(*@$\psi$@*)) 
return c.ID
\end{lstlisting}

\begin{lstlisting}[
language=SQL,
deletekeywords={IDENTITY},
deletekeywords={[2]INT},
morekeywords={clustered},
label={query6},
caption={Response anti-pattern query using the Mutli-Dimensional encoding },escapeinside={(*@}{@*)}]
Match (c:Case) <-[:Event_to_case]- (start:Event{activity:(*@$\phi$@*)})
Match path = (e1:Event{activity:(*@$\psi$@*)}<-[:Directly_follows*]-(start)) 
	where  (start)-[:Directly_follows*]->((*@$\Gamma$@*))-[:Directly_follows*]->(e1) 
	or none (n in nodes(path) where n.activity=(*@$\psi$@*)) 
return c.ID
\end{lstlisting}

\subsubsection{Mapping to Unique Activities (UA) Encoding} \label{sec:unique}
The main purpose for this approach is to simplify the queries and speed up their evaluation by utilizing indexes and skipping the linear scan of the \emph{:Directly\_follows} relation among events. It extends the Baseline approach by adding a \emph{position} attribute to each event node. It creates a compact graph representation model of the event log, resulting in answer compliance queries faster. Table~\ref{tbl:hiringeventlog} has a highlighted column, tagged as \textit{added detail} column, where we assign each event to a position in the case (trace). The utilization of the \emph{position} property simplifies compliance query processing but inherits linear graph size growth concerning log size. This ensures a linear growth with
the size of the set of activity labels by generating separate edges connecting a case node to its corresponding activity node. These edges including various properties like position, timestamp, and resource, representing activity executions within cases. This limits graph growth based on activity labels. 

\begin{figure}[hbt!]
\centering
\includegraphics[width=1\linewidth]{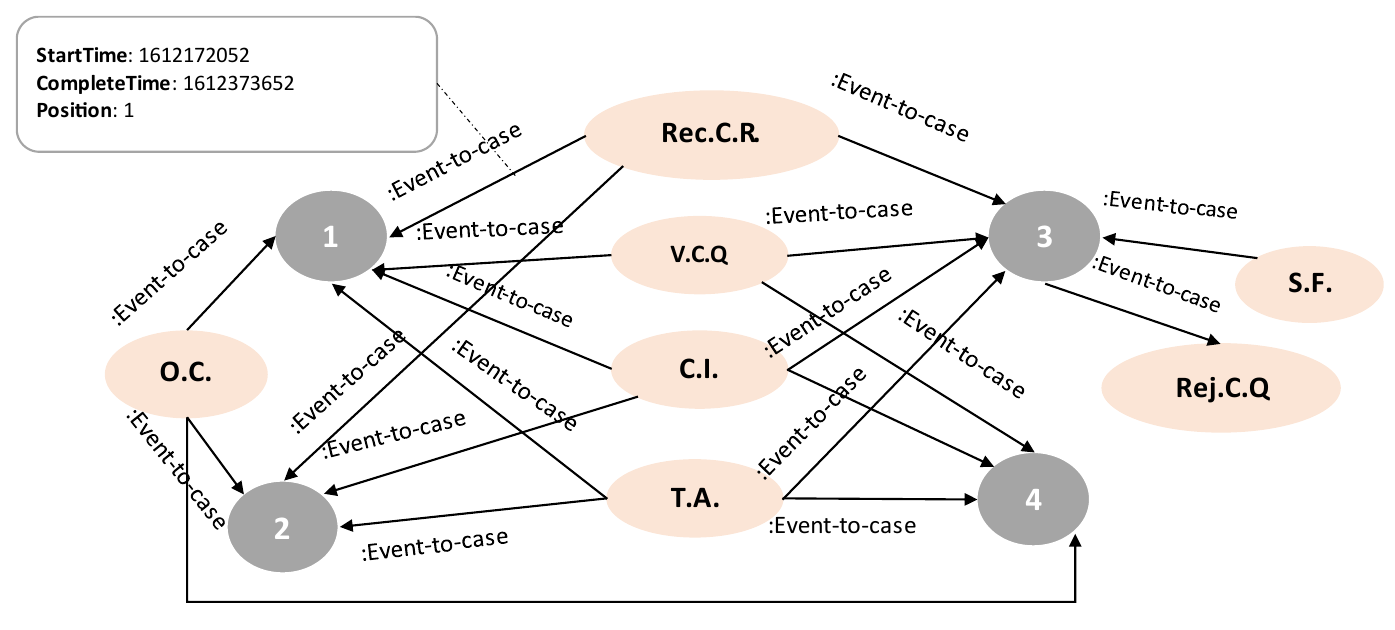}
\caption{UA Encoding of~Table~\ref{tbl:hiringeventlog}}
\label{fig:unique:activities:encodings}    
\end{figure}

Figure~\ref{fig:unique:activities:encodings} shows the graph generated by encoding the log excerpt in Table~\ref{tbl:hiringeventlog} using the unique activities method. Listings \{\ref{query1},\ref{query2},\ref{query3}\} demonstrate the usage of \emph{UA encoding} queries to define compliance patterns under time constraints.

\begin{lstlisting}[
language=SQL,
deletekeywords={IDENTITY},
deletekeywords={[2]INT},
morekeywords={clustered},
label={query1},
caption={Response anti-pattern query using UA encoding},escapeinside={(*@}{@*)}]
Match (c:Case)
where ((c:Case)<-[r1:Event_to_Case]-(a:Event{event:(*@$\phi$@*)}) and (c:case)<-[r2:Event_to_Case]-(b:Event{event:(*@$\psi$@*)}) and (e:Event{event:(*@$\Gamma$@*)})-[r3:Event_to_Case]->(c:Case)) and  (r2.position < r1.position or r3.position > r1.position and r3.position < r2.position)) or not exists((c)<-[:Event_to_Case]-(:Event{event:(*@$\psi$@*)}))
return c.ID
\end{lstlisting} 

\begin{lstlisting}[
language=SQL,
deletekeywords={IDENTITY},
deletekeywords={[2]INT},
morekeywords={clustered},
label={query2},
caption={Response anti-pattern query within time interval (I) using UA encoding},
escapeinside={(*@}{@*)},
linewidth=\textwidth]
Match (a:Event{event:(*@$\phi$@*)})-[r1:Event_to_Case]->(c:Case)<-[r2:Event_to_Case]- (b:Event{event:(*@$\psi$@*)}),(e:Event{event:(*@$\gamma$@*)})-[r3:Event_to_Case]->(c:Case)
where r2.position > r1.position and ((r1.completeTime - r2.startTime) >= I or r3.position > r1.position and r3.position < r2.position)or not exists((c)<-[:Event_to_Case]-(:Event{event:(*@$\psi$@*)}))
return c.ID
\end{lstlisting}

\begin{lstlisting}[
language=SQL,
deletekeywords={IDENTITY},
deletekeywords={[2]INT},
morekeywords={clustered},
label={query3},
caption={Response anti-pattern query after time interval (I) using UA encoding},escapeinside={(*@}{@*)}]
Match (a:Event{event:(*@$\phi$@*)})-[r1:Event_to_Case]->(c:Case)<-[r2:Event_to_Case]- (b:Event{event:(*@$\psi$@*)}),(e:Event{event:(*@$\gamma$@*)})-[r3:Event_to_Case]->(c:Case)
where r2.position > r1.position and ((r2.startTime - r1.completeTime) <= I) or r3.position > r1.position and r3.position < r2.position) or not exists((c)<-[:Event_to_Case]-(:Event{event:(*@$\psi$@*)}))
return c.ID
\end{lstlisting}

\section{Evaluation}\label{sec:eval}

This section assesses the proposed $MTL_f$ mapping to the different data models. The purpose of the evaluation is two-fold. First, we empirically prove that the approach works for the different known log encodings to the best of our knowledge, second we compare the performance of the different encodings against varying log sizes. The metric we report is the execution time of the query(ies) generated to capture the violations, i.e., the anti pattern queries. The real life logs used for evaluation are discussed in Section~\ref{sec:real:logs}. Evaluation setup and other implementation details are presented in Section~\ref{sub:sec:eval:experiment:setting} before we discuss the results in Section~\ref{sub:sec:eval:discuss}.

\subsection{Data sets}
\label{sec:real:logs}

\begin{table}[htb!]
    \caption{Logs Characteristics}
    \label{tbl:char:real:logs}
    \footnotesize
    \centering
    \begin{tabular}{|c|c|c|c|c|} 
    \hline  
    \cellcolor[HTML]{EFEFEF}\textbf{Logs}     &  \textbf{BPIC'12}&  \textbf{BPIC'14}&  \textbf{BPIC'19}&  \textbf{RTFMP}\\ \hline  
         {\cellcolor[HTML]{EFEFEF}\# Traces}&  13087&  41353&  220810&  150370\\ \hline  
 {\cellcolor[HTML]{EFEFEF}\# Events}& 262200& 369485& 979942&561470\\ \hline  
 {\cellcolor[HTML]{EFEFEF}\# Unique Events}& 24& 9& 8&11\\ \hline 
    \end{tabular}

\end{table}

We chose four real-life logs: three obtained from the BPI challenges (BPIC'12 \cite{vanDongen_2012}, BPIC'14\cite{vanDongen_2014}, BPIC'19~\cite{Fahland2021b}) and one additional log, RTFMP~\cite{Reissner2022}. These logs were specifically chosen for their diverse features and distinct characteristics, as summarized in Table~\ref{tbl:char:real:logs}. The rationale behind selecting logs with varied attributes was to ensure a comprehensive evaluation across different data-sets, enabling a robust assessment of our approach in handling diverse real-life scenarios found in business process logs.

\subsection{Implementation Details and Experimental Setup}
\label{sub:sec:eval:experiment:setting}

The approach has been implemented using Python and has been applied across various encoding methods. Given the instantiated $MTL_f$ formula, the event log, the target encoding, our approach generates the respective anti-pattern queries following the templates presented in Section~\ref{sec:proposed:approach}. 

To benchmark the different encoding methods, we have loaded the logs from Table~\ref{tbl:char:real:logs} to the respective data model formats. That is, we stored copies of each log in a relational table and stored the multi-dimensional and the unique activities encodings into separate instances of Neo4J. For the relational databases, we used MS SQL Server and Oracle. We added Oracle to our RDBMSs because MS SQL Server does not support the MR operator. Additionally, the evaluation incorporated the SQL Server Graph Database extension (SGD) version 15.0.2000.


All experiments have been performed using an Intel core i7 processor machine with $16~GB$ RAM. The UA encoding method was implemented utilizing Neo4j version 4.3.1 as the underlying graph database, and Cypher was employed for log querying. The instantiation of Neo4j utilized default settings, comprising a maximum heap size of $2~GB$ and a $512~MB$ page cache size.  was utilized. Oracle and MS SQL Server were started with allocated $1~GB$ RAM and configured with default storage engine settings.

Our evaluation process involved testing our approach across different scenarios, each addressing various compliance patterns. These scenarios were carefully designed to integrate multiple perspectives, incorporating considerations of both data and time. Queries were developed to detect \emph{Response}, \emph{Responded Existence}, \emph{Chain Precedence} and \emph{Precedence} anti-patterns, resulting in the creation of four distinct variants for each pattern. Among these variants, two adhered to temporal constraints by defining upper and lower boundaries for the time window. The other two variants served different purposes: one aimed to exclude specific activities, while the other exclusively focused on the simplest variant of the pattern, e.g., response without exclusion and without a time window.

We have checked the correctness of the results by comparing the reported case identifiers with violations across the ground truth, where we know for each compliance rule what the violating cases are in a log. All the details for the patterns (queries) variants and run details of experiments are available on Github~\footnote{\scriptsize\url{https://github.com/nesmayoussef/CompliancePatterns/}}.

\subsection{Results and Discussion}\label{sub:sec:eval:discuss}

In the first experiment, we run four compliance anti-pattern queries without time condition or any excluded activities for \emph{Response}, \emph{Responded Existence}, \emph{Chain Precedence} and \emph{Precedence} against the respective logs. We have run the queries five times and report the average execution time of these queries for the different encoding methods in Table~\ref{BPI2014-2019:No:time} and Table~\ref{BPI2012:No:time}. Analyzing the comparative performance, it is evident that \emph{UA-Neo4j} consistently demonstrates lower execution times across most patterns compared to other encoding methods. For instance, it shows that \emph{SQL Miner} performed roughly 16x faster than \emph{Declare Analyzer}\footnote{Declare Analyzer is a method for conformance checking that takes log and declare model as input and it generates an output model includes the total number of activations, fulfillments and violations of each constraint in each trace in log.}, indicating a significant speed advantage for \emph{SQL~Miner} in processing the logs. While \emph{UA-Neo4j} emerges as the most efficient performer, it significantly outperforms Declare Analyzer and SQL~Miner. This dataset highlights \emph{UA}'s efficiency in processing these patterns within the specified Neo4j environment. 

\begin{table}[t!]
	\centering
	\caption{Execution time (msec) for the variants of the four queries [No time window]}
	\label{BPI2014-2019:No:time}
	\resizebox{\textwidth}{!}{
		\begin{tabular}{|l|llllllll|}
			\hline
			\rowcolor[HTML]{C0C0C0} 
			\multicolumn{1}{|c|}{\cellcolor[HTML]{C0C0C0}\textbf{Log}}         & \multicolumn{8}{c|}{\cellcolor[HTML]{C0C0C0}\textbf{BPI2019}}      \\ \hline
			\rowcolor[HTML]{C0C0C0} 
			\multicolumn{1}{|c|}{\cellcolor[HTML]{C0C0C0}\textbf{\# of Cases}} & \multicolumn{4}{c||}{\cellcolor[HTML]{C0C0C0}\textbf{25000}}            & \multicolumn{4}{c|}{\cellcolor[HTML]{C0C0C0}\textbf{220810}}          \\ \hline
			\rowcolor[HTML]{EFEFEF} 
			\multicolumn{1}{|c|}{\cellcolor[HTML]{EFEFEF}\textbf{Methods}}     & \multicolumn{1}{l|}{\cellcolor[HTML]{EFEFEF}\textbf{Response}} & \multicolumn{1}{l|}{\cellcolor[HTML]{EFEFEF}\textbf{Precedence}} & \multicolumn{1}{p{2cm}|}{\cellcolor[HTML]{EFEFEF}\textbf{Chain Precedence}} & \multicolumn{1}{p{2cm}||}{\cellcolor[HTML]{EFEFEF}\textbf{Responded-Existence}} & \multicolumn{1}{l|}{\cellcolor[HTML]{EFEFEF}\textbf{Response}} & \multicolumn{1}{l|}{\cellcolor[HTML]{EFEFEF}\textbf{Precedence}} & \multicolumn{1}{p{2cm}|}{\cellcolor[HTML]{EFEFEF}\textbf{Chain Precedence}} & \multicolumn{1}{p{2cm}|}{\cellcolor[HTML]{EFEFEF}\textbf{Responded-Existence}} \\ \hline
			\cellcolor[HTML]{C0C0C0}\textbf{Declare Analyzer} & \multicolumn{1}{l|}{1725}        & \multicolumn{1}{l|}{2125}          & \multicolumn{1}{l|}{2370}             & \multicolumn{1}{l||}{2500} & \multicolumn{1}{l|}{10975}       & \multicolumn{1}{l|}{18011}         & \multicolumn{1}{l|}{16150}            & 14500      \\ \hline
			\cellcolor[HTML]{C0C0C0}\textbf{Match Recognize}           & \multicolumn{1}{l|}{1780}         & \multicolumn{1}{l|}{3171}           & \multicolumn{1}{l|}{1150}              & \multicolumn{1}{l||}{889}  & \multicolumn{1}{l|}{40369}       & \multicolumn{1}{l|}{25864}          & \multicolumn{1}{l|}{2557}              & 16850       \\ \hline
			\cellcolor[HTML]{C0C0C0}\textbf{SQL Miner}           & \multicolumn{1}{l|}{897}         & \multicolumn{1}{l|}{906}           & \multicolumn{1}{l|}{391}              & \multicolumn{1}{l||}{153}  & \multicolumn{1}{l|}{15782}       & \multicolumn{1}{l|}{3222}          & \multicolumn{1}{l|}{984}              & 1428       \\ \hline
			\cellcolor[HTML]{C0C0C0}\textbf{MultiDimensional-SGD}            & \multicolumn{1}{l|}{1593}         & \multicolumn{1}{l|}{2450}          & \multicolumn{1}{l|}{580} & \multicolumn{1}{l||}{420}   & \multicolumn{1}{l|}{16200}            & \multicolumn{1}{l|}{4250}              & \multicolumn{1}{l|}{2106}   &     3210       \\ \hline
			\cellcolor[HTML]{C0C0C0}\textbf{MultiDimensional-Neo4j}            & \multicolumn{1}{l|}{613}         & \multicolumn{1}{l|}{1760}          & \multicolumn{1}{l|}{95} & \multicolumn{1}{l||}{30}   & \multicolumn{1}{l|}{\textbf{\textemdash}}            & \multicolumn{1}{l|}{\textbf{\textemdash}}              & \multicolumn{1}{l|}{\textbf{\textemdash}}   &     \textbf{\textemdash}       \\ \hline
			\cellcolor[HTML]{C0C0C0}\textbf{UA-SGD}            & \multicolumn{1}{l|}{1105}         & \multicolumn{1}{l|}{1210}           & \multicolumn{1}{l|}{380} & \multicolumn{1}{l||}{275}   & \multicolumn{1}{l|}{9760}         & \multicolumn{1}{l|}{3116}           & \multicolumn{1}{l|}{1656}              & 2428        \\ \hline
			\cellcolor[HTML]{C0C0C0}\textbf{UA-Neo4j}            & \multicolumn{1}{l|}{583}         & \multicolumn{1}{l|}{154}           & \multicolumn{1}{l|}{78} & \multicolumn{1}{l||}{40}   & \multicolumn{1}{l|}{144}         & \multicolumn{1}{l|}{145}           & \multicolumn{1}{l|}{450}              & 354        \\ \hline
			\rowcolor[HTML]{C0C0C0} 
			\multicolumn{1}{|c|}{\cellcolor[HTML]{C0C0C0}\textbf{Log}}         & \multicolumn{8}{c|}{\cellcolor[HTML]{C0C0C0}\textbf{BPI2014}}      \\ \hline
			\rowcolor[HTML]{C0C0C0} 
			\multicolumn{1}{|c|}{\cellcolor[HTML]{C0C0C0}\textbf{\# of cases}} & \multicolumn{4}{c||}{\cellcolor[HTML]{C0C0C0}\textbf{15000}}            & \multicolumn{4}{c|}{\cellcolor[HTML]{C0C0C0}\textbf{41373}}           \\ \hline
			\rowcolor[HTML]{EFEFEF} 
			\multicolumn{1}{|c|}{\cellcolor[HTML]{EFEFEF}\textbf{Methods}}     & \multicolumn{1}{l|}{\cellcolor[HTML]{EFEFEF}\textbf{Response}} & \multicolumn{1}{l|}{\cellcolor[HTML]{EFEFEF}\textbf{Precedence}} & \multicolumn{1}{p{2cm}|}{\cellcolor[HTML]{EFEFEF}\textbf{Chain Precedence}} & \multicolumn{1}{p{2cm}||}{\cellcolor[HTML]{EFEFEF}\textbf{Responded-Existence}} & \multicolumn{1}{l|}{\cellcolor[HTML]{EFEFEF}\textbf{Response}} & \multicolumn{1}{l|}{\cellcolor[HTML]{EFEFEF}\textbf{Precedence}} & \multicolumn{1}{p{2cm}|}{\cellcolor[HTML]{EFEFEF}\textbf{Chain Precedence}} & \multicolumn{1}{p{2cm}|}{\cellcolor[HTML]{EFEFEF}\textbf{Responded-Existence}} \\ \hline
			\cellcolor[HTML]{C0C0C0}\textbf{Declare Analyzer} & \multicolumn{1}{l|}{1590}        & \multicolumn{1}{l|}{1498}          & \multicolumn{1}{l|}{310}              & \multicolumn{1}{l||}{370}  & \multicolumn{1}{l|}{3084}        & \multicolumn{1}{l|}{2829}          & \multicolumn{1}{l|}{735}              & 1200       \\ \hline
			\cellcolor[HTML]{C0C0C0}\textbf{Match Recognize}           & \multicolumn{1}{l|}{3612}         & \multicolumn{1}{l|}{3120}           & \multicolumn{1}{l|}{1755}              & \multicolumn{1}{l||}{545}  & \multicolumn{1}{l|}{3149}        & \multicolumn{1}{l|}{2930}          & \multicolumn{1}{l|}{1224}              & 540        \\ \hline
			\cellcolor[HTML]{C0C0C0}\textbf{SQL-Miner}           & \multicolumn{1}{l|}{767}         & \multicolumn{1}{l|}{367}           & \multicolumn{1}{l|}{418}              & \multicolumn{1}{l||}{116}  & \multicolumn{1}{l|}{2249}        & \multicolumn{1}{l|}{1958}          & \multicolumn{1}{l|}{941}              & 385        \\ \hline
			\cellcolor[HTML]{C0C0C0}\textbf{MultiDimensional-SGD}            & \multicolumn{1}{l|}{1411}         & \multicolumn{1}{l|}{2810}          & \multicolumn{1}{l|}{504} & \multicolumn{1}{l||}{466}   & \multicolumn{1}{l|}{3520}            & \multicolumn{1}{l|}{3102}              & \multicolumn{1}{l|}{1260}   &  870          \\ \hline
			\cellcolor[HTML]{C0C0C0}\textbf{MultiDimensional-Neo4j}            & \multicolumn{1}{l|}{166}         & \multicolumn{1}{l|}{1440}          & \multicolumn{1}{l|}{50} & \multicolumn{1}{l||}{63}   & \multicolumn{1}{l|}{\textbf{\textemdash}}            & \multicolumn{1}{l|}{\textbf{\textemdash}}              & \multicolumn{1}{l|}{\textbf{\textemdash}}   &  \textbf{\textemdash}          \\ \hline
			\cellcolor[HTML]{C0C0C0}\textbf{UA-SGD}            & \multicolumn{1}{l|}{1300}         & \multicolumn{1}{l|}{880}           & \multicolumn{1}{l|}{435} & \multicolumn{1}{l||}{272}   & \multicolumn{1}{l|}{2320}         & \multicolumn{1}{l|}{1940}           & \multicolumn{1}{l|}{1199}              & 750  
			\\ \hline
			\cellcolor[HTML]{C0C0C0}\textbf{UA-Neo4j}            & \multicolumn{1}{l|}{133}         & \multicolumn{1}{l|}{109}           & \multicolumn{1}{l|}{50} & \multicolumn{1}{l||}{32}   & \multicolumn{1}{l|}{163}         & \multicolumn{1}{l|}{117}           & \multicolumn{1}{l|}{130}              & 88       \\ \hline
	\end{tabular}}
\end{table}


\begin{table}[t!]
\scriptsize
\centering
\caption{Execution time (msec) for the variants of the four queries for BPI2012 [No time window]}
\label{BPI2012:No:time}
 \resizebox{\textwidth}{!}{
\begin{tabular}{|c|cccc|}
\hline
\rowcolor[HTML]{C0C0C0} 
\textbf{Log} & \multicolumn{4}{c|}{\cellcolor[HTML]{C0C0C0}\textbf{BPI2012}}      \\ \hline
\rowcolor[HTML]{C0C0C0} 
\multicolumn{1}{|l|}{\cellcolor[HTML]{C0C0C0}\textbf{\# of Cases}} & \multicolumn{4}{c|}{\cellcolor[HTML]{C0C0C0}\textbf{13087}}       \\ \hline
\rowcolor[HTML]{EFEFEF} 
\multicolumn{1}{|c|}{\cellcolor[HTML]{C0C0C0}\textbf{Methods}}    & \multicolumn{1}{|c|}{\cellcolor[HTML]{EFEFEF}\textbf{Response}} & \multicolumn{1}{c|}{\cellcolor[HTML]{EFEFEF}\textbf{Precedence}} & \multicolumn{1}{c|}{\cellcolor[HTML]{EFEFEF}\textbf{Chain Precedence}} & \textbf{Responded Existence} \\ \hline
\cellcolor[HTML]{C0C0C0}\textbf{Declare Analyzer}   & \multicolumn{1}{c|}{1938}          & \multicolumn{1}{c|}{1697}            & \multicolumn{1}{c|}{407}     & 385            \\ \hline
\cellcolor[HTML]{C0C0C0}\textbf{Match Recognize}           & \multicolumn{1}{c|}{4780}          & \multicolumn{1}{c|}{2390}             & \multicolumn{1}{c|}{797}     & 350            \\ \hline
\cellcolor[HTML]{C0C0C0}\textbf{SQL Miner}           & \multicolumn{1}{c|}{1656}          & \multicolumn{1}{c|}{747}             & \multicolumn{1}{c|}{275}     & 109            \\ \hline
\cellcolor[HTML]{C0C0C0}\textbf{MultiDimensional-SGD}            & \multicolumn{1}{c|}{670}           & \multicolumn{1}{c|}{1180}            & \multicolumn{1}{c|}{230}      & 138             \\ \hline
\cellcolor[HTML]{C0C0C0}\textbf{MultiDimensional-Neo4j}            & \multicolumn{1}{c|}{203}           & \multicolumn{1}{c|}{1750}            & \multicolumn{1}{c|}{66}      & 38             \\ \hline
\cellcolor[HTML]{C0C0C0}\textbf{UA-SGD}            & \multicolumn{1}{c|}{402}           & \multicolumn{1}{c|}{712}              & \multicolumn{1}{c|}{184}      & 124             \\ \hline
\cellcolor[HTML]{C0C0C0}\textbf{UA-Neo4j}            & \multicolumn{1}{c|}{128}           & \multicolumn{1}{c|}{54}              & \multicolumn{1}{c|}{28}      & 10             \\ \hline
\end{tabular}
}
\end{table}


In the second experiment, anti-pattern queries were executed across various encoding methods, excluding Declare Analyzer as it did not support the ``exclude'' operation. As illustrated in Table~\ref{BPI2019-2014:Exclude:with:time} and Table~\ref{BPI2012:Exclude:With:Time}, the comparison highlights a noteworthy observation with the \emph{UA-Neo4j} method exhibiting better performance, approximately 8 times quicker than \emph{UA-SGD}. Moreover, \emph{UA-Neo4j} is notably faster than \emph{SQL Miner} by 14x because \emph{SQL Miner} employs nested queries and self-joins. These queries are more expensive to evaluate, imposing extra tasks on the query engine during data retrieval.

\begin{table}[t!]
\centering
\caption{Execution time (msec) for the variants of the Exclude queries [With time window]}
\label{BPI2019-2014:Exclude:with:time}
\resizebox{\textwidth}{!}{
\begin{tabular}{|c|cccccccc|}
\hline
\rowcolor[HTML]{C0C0C0} 
\textbf{Logs} & \multicolumn{8}{c|}{\cellcolor[HTML]{C0C0C0}\textbf{BPI2019}}      \\ \hline
\rowcolor[HTML]{C0C0C0} 
\textbf{\# of Cases}        & \multicolumn{4}{c||}{\cellcolor[HTML]{C0C0C0}\textbf{25000}}   & \multicolumn{4}{c|}{\cellcolor[HTML]{C0C0C0}\textbf{220810}}             \\ \hline
\rowcolor[HTML]{EFEFEF} 
\multicolumn{1}{|c|}{\cellcolor[HTML]{EFEFEF}\textbf{Methods}}     & \multicolumn{1}{l|}{\cellcolor[HTML]{EFEFEF}\textbf{Response}} & \multicolumn{1}{l|}{\cellcolor[HTML]{EFEFEF}\textbf{Precedence}} & \multicolumn{1}{p{2cm}|}{\cellcolor[HTML]{EFEFEF}\textbf{Chain Precedence}} & \multicolumn{1}{p{2cm}||}{\cellcolor[HTML]{EFEFEF}\textbf{Responded-Existence}} & \multicolumn{1}{l|}{\cellcolor[HTML]{EFEFEF}\textbf{Response}} & \multicolumn{1}{l|}{\cellcolor[HTML]{EFEFEF}\textbf{Precedence}} & \multicolumn{1}{p{2cm}|}{\cellcolor[HTML]{EFEFEF}\textbf{Chain Precedence}} & \multicolumn{1}{p{2cm}|}{\cellcolor[HTML]{EFEFEF}\textbf{Responded-Existence}} \\ \hline

\multicolumn{1}{|l|}{\cellcolor[HTML]{C0C0C0}Match Recognize} & \multicolumn{1}{c|}{3810}          & \multicolumn{1}{c|}{2128}             & \multicolumn{1}{c|}{1560}  & \multicolumn{1}{c||}{2750}   & \multicolumn{1}{c|}{60320}         & \multicolumn{1}{c|}{20578}            & \multicolumn{1}{c|}{13524} & 22950      \\ \hline
\multicolumn{1}{|l|}{\cellcolor[HTML]{C0C0C0}SQL Miner} & \multicolumn{1}{c|}{2310}          & \multicolumn{1}{c|}{553}             & \multicolumn{1}{c|}{687}  & \multicolumn{1}{c||}{1624}   & \multicolumn{1}{c|}{20415}         & \multicolumn{1}{c|}{4885}            & \multicolumn{1}{c|}{6075} & 14341      \\ \hline
\multicolumn{1}{|l|}{\cellcolor[HTML]{C0C0C0}MultiDimensional-Neo4j}  & \multicolumn{1}{c|}{289}           & \multicolumn{1}{c|}{182}             & \multicolumn{1}{c|}{230}  & \multicolumn{1}{c||}{420}    & \multicolumn{1}{c|}{\textbf{\textemdash}}              & \multicolumn{1}{c|}{\textbf{\textemdash}}  & \multicolumn{1}{c|}{\textbf{\textemdash}}     &   \textbf{\textemdash}         \\ \hline
\multicolumn{1}{|l|}{\cellcolor[HTML]{C0C0C0}MultiDimensional-SGD}    & \multicolumn{1}{c|}{725}           & \multicolumn{1}{c|}{1013}            & \multicolumn{1}{c|}{860}  & \multicolumn{1}{c||}{1226}   & \multicolumn{1}{c|}{6412}          & \multicolumn{1}{c|}{8947}            & \multicolumn{1}{c|}{7529} & 10950      \\ \hline
\multicolumn{1}{|l|}{\cellcolor[HTML]{C0C0C0}UA-SGD}    & \multicolumn{1}{c|}{506}           & \multicolumn{1}{c|}{692}             & \multicolumn{1}{c|}{421}  & \multicolumn{1}{c||}{670}    & \multicolumn{1}{c|}{4480}          & \multicolumn{1}{c|}{6100}            & \multicolumn{1}{c|}{3711} & 5922       \\ \hline
\multicolumn{1}{|l|}{\cellcolor[HTML]{C0C0C0}UA-Neo4j}  & \multicolumn{1}{c|}{162}           & \multicolumn{1}{c|}{96}              & \multicolumn{1}{c|}{119}  & \multicolumn{1}{c||}{193}    & \multicolumn{1}{c|}{1434}          & \multicolumn{1}{c|}{847}             & \multicolumn{1}{c|}{1047} & 1620       \\ \hline
\rowcolor[HTML]{C0C0C0} 
\textbf{Log}  & \multicolumn{8}{c|}{\cellcolor[HTML]{C0C0C0}\textbf{BPI2014}}      \\ \hline
\rowcolor[HTML]{C0C0C0} 
\textbf{\# of Cases}        & \multicolumn{4}{c||}{\cellcolor[HTML]{C0C0C0}\textbf{15000}}   & \multicolumn{4}{c|}{\cellcolor[HTML]{C0C0C0}\textbf{41373}}              \\ \hline
\rowcolor[HTML]{EFEFEF} 
\multicolumn{1}{|c|}{\cellcolor[HTML]{EFEFEF}\textbf{Methods}}     & \multicolumn{1}{l|}{\cellcolor[HTML]{EFEFEF}\textbf{Response}} & \multicolumn{1}{l|}{\cellcolor[HTML]{EFEFEF}\textbf{Precedence}} & \multicolumn{1}{p{2cm}|}{\cellcolor[HTML]{EFEFEF}\textbf{Chain Precedence}} & \multicolumn{1}{p{2cm}||}{\cellcolor[HTML]{EFEFEF}\textbf{Responded-Existence}} & \multicolumn{1}{l|}{\cellcolor[HTML]{EFEFEF}\textbf{Response}} & \multicolumn{1}{l|}{\cellcolor[HTML]{EFEFEF}\textbf{Precedence}} & \multicolumn{1}{p{2cm}|}{\cellcolor[HTML]{EFEFEF}\textbf{Chain Precedence}} & \multicolumn{1}{p{2cm}|}{\cellcolor[HTML]{EFEFEF}\textbf{Responded-Existence}} \\ \hline

\cellcolor[HTML]{C0C0C0}Match Recognize         & \multicolumn{1}{c|}{4320}          & \multicolumn{1}{c|}{1228}             & \multicolumn{1}{c|}{1445}  & \multicolumn{1}{c||}{2800}    & \multicolumn{1}{c|}{6520}          & \multicolumn{1}{c|}{2287}             & \multicolumn{1}{c|}{3641} & 5420       \\ \hline
\cellcolor[HTML]{C0C0C0}SQL Miner         & \multicolumn{1}{c|}{1386}          & \multicolumn{1}{c|}{332}             & \multicolumn{1}{c|}{413}  & \multicolumn{1}{c||}{975}    & \multicolumn{1}{c|}{3825}          & \multicolumn{1}{c|}{915}             & \multicolumn{1}{c|}{1138} & 2587       \\ \hline
\cellcolor[HTML]{C0C0C0}MultiDimensional-Neo4j          & \multicolumn{1}{c|}{174}           & \multicolumn{1}{c|}{110}             & \multicolumn{1}{c|}{138}  & \multicolumn{1}{c||}{252}    & \multicolumn{1}{c|}{\textbf{\textemdash}}              & \multicolumn{1}{c|}{\textbf{\textemdash}}  & \multicolumn{1}{c|}{\textbf{\textemdash}}     &    \textbf{\textemdash}        \\ \hline
\cellcolor[HTML]{C0C0C0}MultiDimensional-SGD            & \multicolumn{1}{c|}{435}           & \multicolumn{1}{c|}{608}             & \multicolumn{1}{c|}{516}  & \multicolumn{1}{c||}{736}    & \multicolumn{1}{c|}{1201}          & \multicolumn{1}{c|}{1676}            & \multicolumn{1}{c|}{1410} & 2052       \\ \hline
\cellcolor[HTML]{C0C0C0}UA-SGD            & \multicolumn{1}{c|}{304}           & \multicolumn{1}{c|}{416}             & \multicolumn{1}{c|}{253}  & \multicolumn{1}{c||}{402}    & \multicolumn{1}{c|}{840}           & \multicolumn{1}{c|}{1143}            & \multicolumn{1}{c|}{696}  & 1107       \\ \hline
\cellcolor[HTML]{C0C0C0}UA-Neo4j          & \multicolumn{1}{c|}{98}            & \multicolumn{1}{c|}{58}              & \multicolumn{1}{c|}{72}   & \multicolumn{1}{c||}{116}    & \multicolumn{1}{c|}{269}           & \multicolumn{1}{c|}{159}             & \multicolumn{1}{c|}{197}  & 305        \\ \hline
\end{tabular}}
\end{table}


\begin{table}[t!]
\centering
\scriptsize
\caption{Execution time (msec) for the variants of the Exclude queries for BPI2012 [With time window]}
\label{BPI2012:Exclude:With:Time}
\resizebox{\textwidth}{!}{
\begin{tabular}{|
>{\columncolor[HTML]{C0C0C0}}c |cccc|}
\hline
\rowcolor[HTML]{C0C0C0} 
\textbf{Log} & \multicolumn{4}{c|}{\cellcolor[HTML]{C0C0C0}\textbf{BPI2012}}      \\ \hline
\textbf{\# of Cases}              & \multicolumn{4}{c|}{\cellcolor[HTML]{C0C0C0} \textbf{13087}}              \\ \hline
\cellcolor[HTML]{C0C0C0}\textbf{Methods} & \multicolumn{1}{c|}{\cellcolor[HTML]{EFEFEF}\textbf{Response}} & \multicolumn{1}{c|}{\cellcolor[HTML]{EFEFEF}\textbf{Precedence}} & \multicolumn{1}{c|}{\cellcolor[HTML]{EFEFEF}\textbf{Chain Preced.}} & \cellcolor[HTML]{EFEFEF}\textbf{Responded-Existence} \\ \hline
\textbf{Match Recognize}  & \multicolumn{1}{c|}{3514}          & \multicolumn{1}{c|}{1015}             & \multicolumn{1}{c|}{1152}  & 2320        \\ \hline
\textbf{SQL Miner}  & \multicolumn{1}{c|}{1210}          & \multicolumn{1}{c|}{290}             & \multicolumn{1}{c|}{360}  & 850        \\ \hline
\textbf{MultiDimensional-Neo4j}   & \multicolumn{1}{c|}{150}           & \multicolumn{1}{c|}{95}              & \multicolumn{1}{c|}{120}  & 220        \\ \hline
\textbf{MultiDimensional-SGD}     & \multicolumn{1}{c|}{380}           & \multicolumn{1}{c|}{530}             & \multicolumn{1}{c|}{450}  & 642        \\ \hline
\textbf{UA-SGD}     & \multicolumn{1}{c|}{265}           & \multicolumn{1}{c|}{362}             & \multicolumn{1}{c|}{220}  & 351        \\ \hline
\textbf{UA-Neo4j}   & \multicolumn{1}{c|}{85}            & \multicolumn{1}{c|}{50}              & \multicolumn{1}{c|}{62}   & 96         \\ \hline
\end{tabular}
}
\end{table}

\break
In a comprehensive assessment across various anti-pattern queries, the utilization of graph-based encoding for event logs consistently surpasses the performance of relational database encoding. This supports the current trend favoring graph databases for process analytics~\cite{multiDimlEventGrahDB21}. Furthermore, the $UA$ encoding method demonstrates notable enhancements in both query time and storage space when compared to the multi-dimensional graph encoding method and other relational encoding methods.

\section{Conclusion and Future Work}\label{sec:conclusion:future}

In this paper, we aim to provide a comprehensive approach to formally represent compliance rules. Our formalism is presented using Metric Temporal Logic over finite traces $(MTL_f)$; which generalizes Linear Temporal Logic over finite traces $(LTL_f)$ to cover compliance patterns with implicit and explicit time constraints.

To perform conformance checking, we map the negated $MTL_f$ formulas, i.e. the anti-patterns, to various query languages supported by the target encoding of the event logs. In general, we cover relational and graph encodings. The purpose of these queries is to identify cases that exhibit violations of the compliance rules.

All mappings were evaluated over four real-life event logs. The results of mapping to Unique Activities $(UA)$ show superior results in comparison to the other encoding approaches in terms of query execution time.

As a future work, we intend to extend the evaluation of our approach across a set of data-aware and resource-aware compliance patterns. Moreover, we intend to stress-test the different encodings to get an understanding of the processing limits of each encoding.


\bibliographystyle{elsarticle-num}

\end{document}